\documentclass[lettersize,journal]{IEEEtran}
\usepackage{amsmath,amsfonts}
\usepackage{algorithmic}
\usepackage{algorithm}
\usepackage{array}
\usepackage{textcomp}
\usepackage{stfloats}
\usepackage{url}
\usepackage{verbatim}
\usepackage{graphicx}
\usepackage{cite}
\usepackage{subfigure, multirow, booktabs, color}
\usepackage{ragged2e}
\renewcommand{\justify}{\leftskip=0pt \rightskip=0pt plus 0cm}
\begin{document}

\title{Advancing Learned Video Compression with In-loop Frame Prediction}

\author{Ren Yang, Radu Timofte,~\IEEEmembership{Member,~IEEE,} Luc Van Gool,~\IEEEmembership{Member,~IEEE}
        % <-this % stops a space
\thanks{Ren Yang is with the Computer Vision Lab, D-ITET, ETH Zurich, 8092 Zurich, Switzerland. E-mail: ren.yang@vision.ee.ethz.ch}
\thanks{Radu Timofte is with the Julius Maximilian University of W\"urzburg, 97070 W\"urzburg, Germany, and ETH Zurich, 8092 Zurich, Switzerland. E-mail: radu.timofte@uni-wuerzburg.de}
\thanks{Luc Van Gool is with the Computer Vision Lab, D-ITET, ETH Zurich, 8092 Zurich, Switzerland, and KU Leuven, 3000 Leuven, Belgium. E-mail: vangool@vision.ee.ethz.ch}
\thanks{This work was partly supported by ETH Zurich General Fund and Humboldt Foundation.}

}% <-this % stops a space


\IEEEpubid{0000--0000/00\$00.00~\copyright~2021 IEEE}

\maketitle

\begin{abstract}
Recent years have witnessed an increasing interest in end-to-end learned video compression. Most previous works explore temporal redundancy by detecting and compressing a motion map to warp the reference frame towards the target frame. Yet, it failed to adequately take advantage of the historical priors in the sequential reference frames. In this paper, we propose an Advanced Learned Video Compression (ALVC) approach with the in-loop frame prediction module, which is able to effectively predict the target frame from the previously compressed frames, \textit{without consuming any bit-rate}. The predicted frame can serve as a better reference than the previously compressed frame, and therefore it benefits the compression performance. The proposed in-loop prediction module is a part of the end-to-end video compression and is jointly optimized in the whole framework. We propose the recurrent and the bi-directional in-loop prediction modules for compressing P-frames and B-frames, respectively. The experiments show the state-of-the-art performance of our ALVC approach in learned video compression. We also outperform the default hierarchical B mode of x265 in terms of PSNR and beat the slowest mode of the SSIM-tuned x265 on MS-SSIM. The project page: \url{https://github.com/RenYang-home/ALVC}. 

\end{abstract}


\section{Introduction}

\IEEEPARstart{V}{ideo} steaming over the Internet becomes more and more popular, and the demands of transmitting high quality and high resolution videos are also rapidly increasing. Video compression plays an important role. During the past decades, plenty of algorithms have been standardized, such as H.264~\cite{wiegand2003overview}, H.265~\cite{sullivan2012overview}, etc. Recently, inspired by the success of end-to-end learned image compression, a great number of learned video compression methods~\cite{xu2020learned,lu2021deep} have been proposed.

As before with handcrafted algorithms, it is essential for learned video compression methods to explore the correlation among video frames to reduce the temporal redundancy, thus reducing the bit-rate. Most previous works, e.g., DVC~\cite{lu2019dvc,lu2020end}, HLVC~\cite{yang2020learning}, Agustsson~et al.~\cite{agustsson2020scale}, RLVC~\cite{yang2020recurrent} and FVC~\cite{hu2021fvc}, detect the temporal motion map to compensate the previously compressed frames. This effectively reduces the temporal redundancy. Nevertheless, the motion map consumes bit-rate, and they fail to make use of the historical motion prior, which may be utilized to predict the target frame without consuming any bit-rate. Therefore, incorporating an in-loop frame prediction module may effectively advance the performance of learned video compression.

\begin{figure*}[!t]
\centering
\includegraphics[width=.9\linewidth]{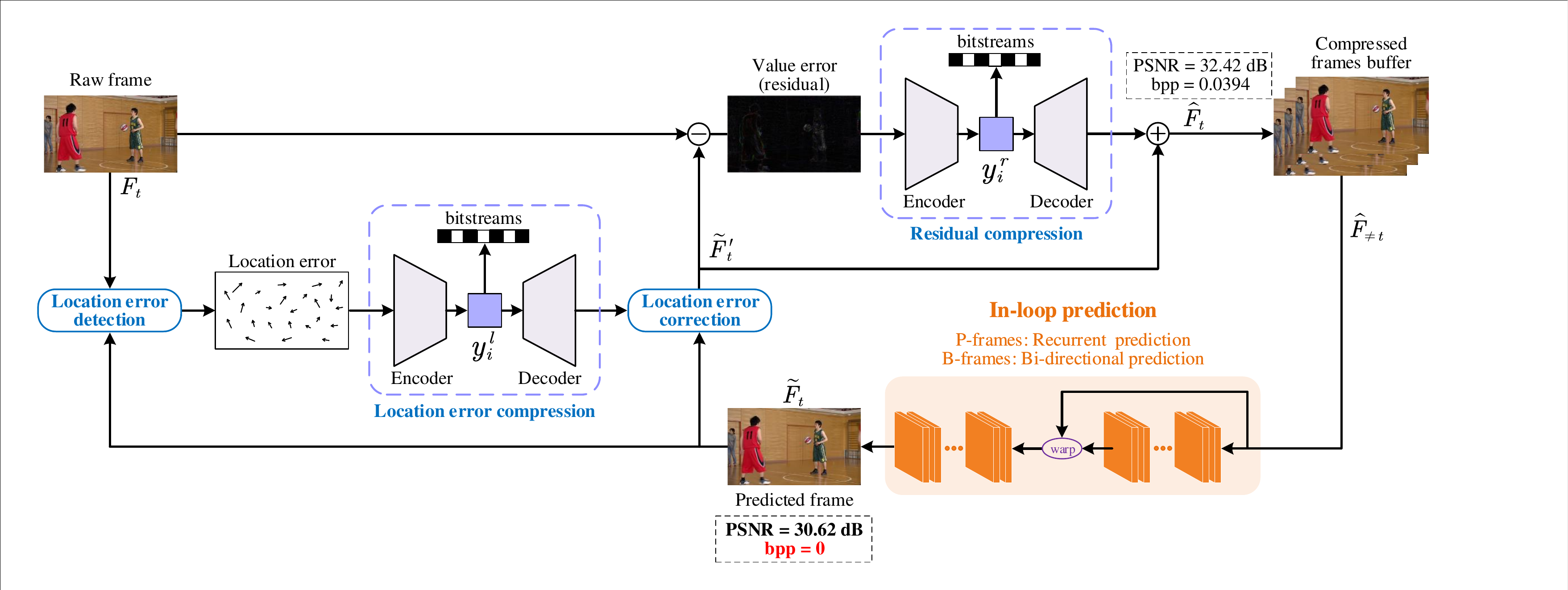}
\caption{High-level architecture of the proposed ALVC approach.} \label{fig:1}
\end{figure*}

{In this paper, we propose an Advanced Learned Video Compression (ALVC) approach with in-loop frame prediction. Fig.~~\ref{fig:1} illustrates the high-level architecture of the proposed ALVC approach. The proposed in-loop frame prediction module is a part of the video compression loop, taking as inputs the previously compressed frames and generating the prediction of the current frame without consuming any bit-rate. Given this predicted frame, we then detect the error between the pixel locations in the raw and predicted frame, and then compress the location error by an auto-encoder and use the compressed location error to correct the pixel locations in the predicted frame. Finally, another auto-encoder is employed to compress the residual.} In Fig.~\ref{fig:1}, we use the third frame in \textit{BasketballPass} as an example. The proposed in-loop prediction module generates the predicted frame with PSNR of 30.62 dB to the target frame with \textbf{\textit{zero bit-rate}} consumption, and only the difference between the predicted and raw frames needs to be compressed into bitstreams, i.e., using 0.0394 bpp to improve the PSNR by 1.8 dB to reach 32.42 dB as the compressed frame. As a result, our ALVC approach achieves better performance than the existing learned compression approaches.
\IEEEpubidadjcol

It is worth pointing out that the proposed in-loop prediction for learned video compression is different from video extrapolation and interpolation. In video extra/interpolation, the prior frames are fixed, and thus the prediction error increases along the distance from the existing frames. Therefore the long-distance (multi-frame) prediction is difficult. However, in our ALVC with in-loop prediction, the compressed frame is recursively added into the buffer and serves as historical prior. Hence, we are able to predict the target frame in a recursive manner, and achieve accurate prediction along time steps. More importantly, the proposed prediction module is a part of the compression loop, and it is jointly optimized with the video compression network in an end-to-end manner.

The contributions of this paper can be summarized as follows:
\begin{itemize}
   \item We propose a novel learned video compression architecture with in-loop frame prediction. 
   \item We propose the in-loop recurrent uni-directional frame prediction module for compressing P-frames, and propose the in-loop bi-directional frame prediction module for compressing B-frames.
   \item The experiments show that the proposed ALVC approach achieves state-of-the-art performance in learned video compression.
   \item The ablation experiments verify the effectiveness of the proposed in-loop prediction and the flexibility for various GOP structures.
  
\end{itemize}

\section{Related work}

\textbf{Learned image compression.} 
For image compression beyond the handcrafted standards (JPEG~\cite{wallace1992jpeg}, BPG~\cite{BPG} and VVC~\cite{bross2021overview}), plenty of works~\cite{Toderici2016Variable,toderici2017full,agustsson2017soft,theis2017lossy,balle2017end,balle2018variational,minnen2018joint,mentzer2018conditional,li2018learning,johnston2018improved,lee2019context,Hu2020Coarse,ma2020end,cheng2020learned,he2021checkerboard,xie2021enhanced,strumpler2021implicit} have focused on end-to-end learned image compression. 
At the beginning, Ball{\'e}~et al.~\cite{balle2017end,balle2018variational} proposed utilizing a variational auto-encoder for deep image compression with factorized~\cite{balle2017end} and hyperprior~\cite{balle2018variational} entropy models. Then, Minnen~et al.~\cite{minnen2018joint} and Lee~et al.~\cite{lee2019context} proposed auto-regressive entropy models to improve the compression efficiency. Later, the coarse-to-fine hyperprior~\cite{Hu2020Coarse} entropy model and the wavelet-like deep transformer~\cite{ma2020end} were designed to further advance the rate-distortion performance, and successfully outperform the image coding standard BPG~\cite{BPG}. Most recently, thanks to advanced coding strategies, such as the invertible auto-encoder~\cite{xie2021enhanced}, the Gaussian mixture model and the attention module~\cite{cheng2020learned}, the latest learned image methods successfully beat the last coding standard VVC.

{\textbf{Handcrafted video compression.} In the past decades, several handcrafted video compression algorithms have been standardized, such as MPEG~\cite{le1992mpeg}, H.264~\cite{wiegand2003overview}, HEVC~\cite{sullivan2012overview} and VVC~\cite{bross2021overview}. In these standards, VVC~\cite{bross2021overview} is the latest video coding standard and performs best among the handcrafted video compression methods. Besides, there are a great number of works that aim at improving the performance the handcrafted video compression algorithms. In these works, the motion refinement technology is most related to our work. For example, Liu~et al.~\cite{liu2010motion} proposed a Multi-Resolution Motion Refinement
(MRMR) scheme at the decoder side to refine motion estimation from the previously decoded data. Youn~et al.~\cite{youn1999motion} proposed a fast-search adaptive motion vector refinement method to refine the motion vectors instead of directly using the incoming motion vectors in transcoding.}

\textbf{Learned video compression.} Inspired by the success of learned image compression, many end-to-end learned video compression methods~\cite{wu2018video,lu2019dvc,lu2020end,yang2020opendvc,cheng2019learning,habibian2019video,djelouah2019neural,liu2019learned,yang2020learning,lin2020m,agustsson2020scale,golinski2020feedback,lu2020content,hu2020improving,yang2020recurrent,hu2021fvc,yang2021perceptual,liu2022end} have been proposed. 
% In 2018, Wu~et al.~\cite{wu2018video} proposed predicting frames by interpolation from reference frames, and compressing residual by the image compression model~\cite{toderici2017full}. 
For instance, Lu~et al.~\cite{lu2019dvc,lu2020end} proposed the first end-to-end Deep Video Compression (DVC) approach. Then, Liu~et al.~\cite{liu2019learned} proposed a one-stage flow for motion compensation. Moreover, the content adaptive and error propagation aware model~\cite{lu2020content} and the resolution-adaptive flow coding~\cite{hu2020improving} strategies were employed to improve compression efficiency. Lin~et al.~\cite{lin2020m} extended the number of reference frames. Meanwhile, Yang~et al.~\cite{yang2020learning} employed hierarchical quality layers, and Agustsson~et al.~\cite{agustsson2020scale} proposed  scale-space flow for learned video compression. Later, Golinkski~et al.~\cite{golinski2020feedback} and Yang~et al.~\cite{yang2020recurrent} proposed recurrent frameworks to make better use of the temporal information. Most recently, the FVC method~\cite{hu2021fvc} was proposed to perform video compression in the feature domain. Liu~et  al.~\cite{liu2022end} proposed the hybrid motion compensation with compound spatio-temporal representation in end-to-end learned video compression, and
{Li~et al.~\cite{li2021deep} proposed the DCVC method, which uses  feature domain context for temporally conditional coding.} 

{Different from these existing works, this paper proposes a neural video compression approach with in-loop frame prediction networks, which is able to effectively predict the target frame from the previously compressed frames, without consuming any bit-rate. The predicted frame can serve as a better reference than the previously compressed frame, and therefore it benefits the compression performance.}

\textbf{Frame prediction.} Frame prediction has been studied to increase the frame-rate (interpolation) and to predict future frames (extrapolation). For example, Niklaus~et al. proposed Adaptive Convolution (AdaConv)~\cite{niklaus2017video} and Separable Convolution (SepConv)~\cite{niklaus2017sep} methods for video interpolation. Meanwhile, Liu~et al.~\cite{liu2017video} proposed predicting the 3D voxel flow to synthesize intermediate or future frames. Moreover, the PhaseNet~\cite{meyer2018phasenet} and Super-SloMo~\cite{jiang2018super} methods were proposed to handle challenging scenarios (large motion, etc.) and multi-frame interpolation, respectively.
Recently, Quadratic Video Interpolation (QVI)~\cite{xu2019quadratic,liu2020enhanced} provided a higher-order motion model, using acceleration for a more precise interpolation. For video extrapolation, Finn~et al.~\cite{finn2016unsupervised} constructed a ConvLSTM-based method to predict future frames. Later, PredRNN~(++)~\cite{wang2017predrnn,wang2018predrnn++} was proposed to memorize spatial appearances and temporal variations for the generation of future frames. Moreover, the 3D-LSTM-based method~\cite{wang2018eidetic},  spatial-temporal multi-frequency analysis~\cite{jin2020exploring} and  convolutional tensor-train decomposition~\cite{su2020convolutional} were introduced into video extrapolation. Most recently, the LMC-Memory method~\cite{lee2021video} employs memory alignment learning to store long-term motion contexts and match them with sequences including limited dynamics. 

{\textbf{Deep frame prediction in hybrid video compression.}
Deep frame prediction generates video frames without consuming bit-rate. Therefore, incorporating frame prediction in video compression is able to advance the rate-distortion performance. Choi~et al.~\cite{choi2019deep} proposed a deep frame prediction network inspired by \cite{niklaus2017sep} in HEVC~\cite{sullivan2012overview}, reducing the bit-rate of HEVC by 2.3\% to 4.4\%. Then, Xia~et al.~\cite{xia2019deep} achieved an average bits reduction of 5.7\% by proposing the Multiscale Adaptive
Separable Convolutional Neural Network (MASCNN) for deep frame prediction. Later, the affine transformation-based deep frame prediction method was proposed in Choi~et al.~\cite{choi2021affine} to improve the efficiency of HEVC. It further advances the rate-distortion performance with fewer parameters than the previous works \cite{choi2019deep} and \cite{xia2019deep}. Most recently, Jin~et al.~\cite{jin2021deep} proposed a deep affine motion compensation network to deal with the deformable motion compensation and applied the proposed method in VVC~\cite{bross2021overview}.}

\begin{figure*}[!t]
\centering
\includegraphics[width=.9\linewidth]{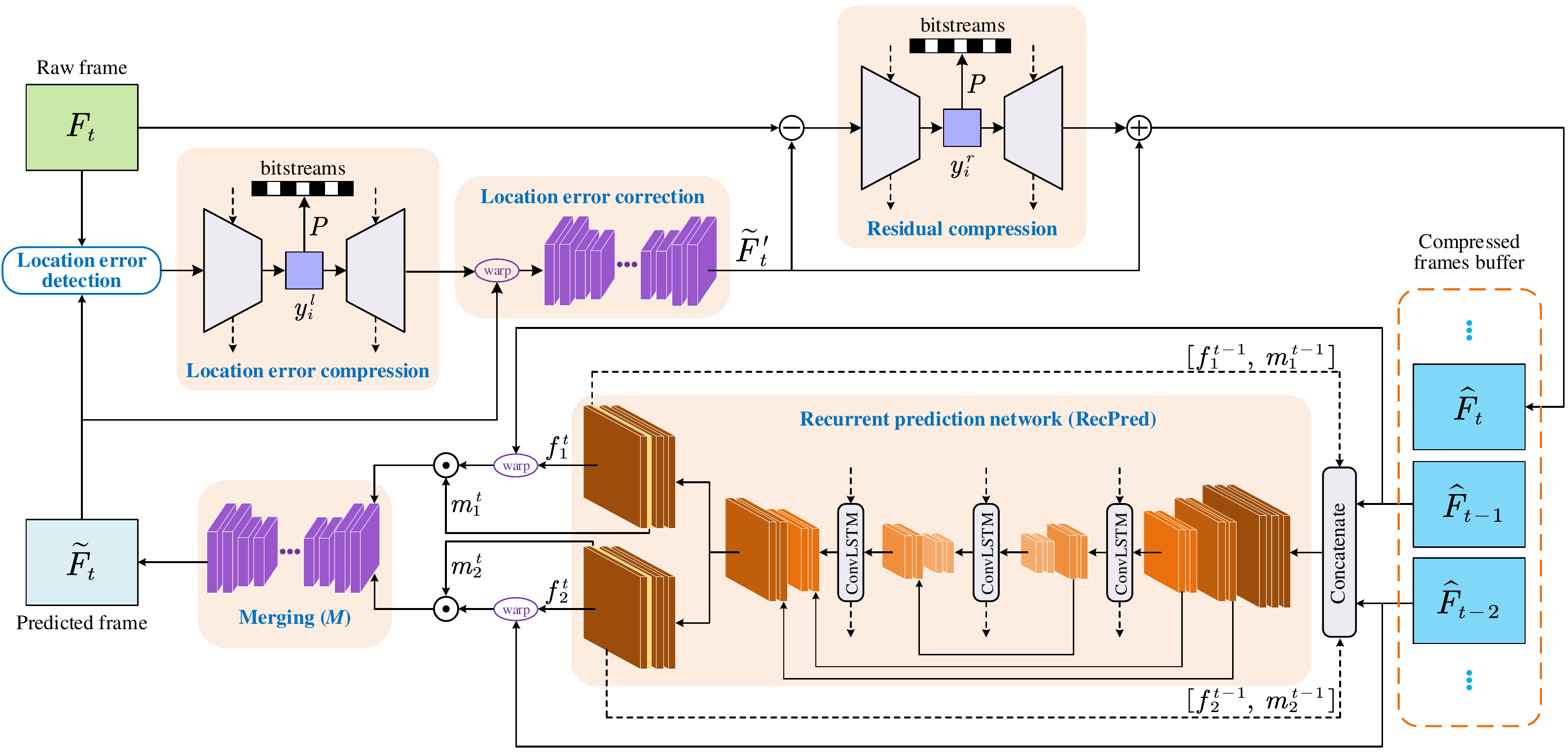}
\caption{The proposed ALVC for P-frames with recurrent in-loop prediction network.} \label{fig:2}
\end{figure*}

{Different above methods, this paper proposes incorporating deep frame prediction as an in-loop component of the end-to-end optimized video compression framework. As such, the whole framework is able to be jointly optimized in an end-to-end manner.} 

\section{Proposed ALVC approach}\label{method}

Fig.~\ref{fig:1} illustrates the high-level framework of the proposed ALVC approach. At time step $t$, we define the raw frame as $F_t$ and the compressed frame as $\hat F_t$. As shown in Fig.~\ref{fig:1}, we propose the ALVC approach with the novel in-loop prediction modules, which predict the target frame from the previously compressed frames $\hat F_{\neq t}$. Due to the high correlation and the temporal coherence of video frames, the target frame can be accurately predicted. Most previous works~\cite{lu2019dvc,yang2020learning,lin2020m,yang2020recurrent,hu2021fvc} compress the motion between the target and previous frames into a bitstream, while the in-loop prediction of our ALVC approach compensates most of the temporal motion with \textbf{\textit{zero bit-rate}}, and only the error between the predicted frame $\tilde F_t$ and the raw frame $F_t$ needs to be compressed into bitstreams.

We decompose the prediction error into location error and value error. The location error is defined as the error between the coordinates of the predicted pixels and groundtruth frames (because of imperfect temporal prediction), and the value error indicates the residual between the predicted pixel values and their groundtruth. In our ALVC approach, we borrow SPyNet~\cite{ranjan2017optical}, which is originally designed for optical flow, to detect the location error, and employ an auto-encoder to compress the location error. Note that, since the proposed in-loop prediction network predicts the consecutive motion from previously compressed frames, the location error in ALVC is expected to be smaller than the motion between the previous and the target frames. Thus, the bit-rate used to correct the location error is less than that to compensate the temporal motion. Given the compressed location error, we correct the location error by warping the predicted frame $\tilde F_t$ and then feed it into a convolutional network to reduce the warping artifacts and increase the nonlinearity of the proposed framework (as shown in Fig.~\ref{fig:2}). The predicted frame after location error correction is defined as $\tilde F'_t$. Afterwards, the value error (residual) is calculated as the difference between the raw frame $F_t$ and $\tilde F'_t$, and then another auto-encoder is utilized to compress the residual. Finally, the compressed frames are generated by adding the compressed residual to $\tilde F'_t$. After the compression of $F_t$, the compressed frame $\hat F_t$ will be a part of the historical prior for predicting future frames. This way, the proposed prediction network becomes an in-loop module of the end-to-end learned video compression framework. 
In ALVC, we propose different in-loop prediction networks for compressing the P-frames and B-frames, which are introduced next.

\subsection{P-frame: ALVC with recurrent in-loop prediction}

In video compression in IPPP mode, the P-frames are consecutively compressed in a uni-directional way. Therefore, to compress P-frames, we propose ALVC with a Recurrent in-loop Prediction (RecPred) network. 

The architecture is shown in Fig.~\ref{fig:2}. 
We use U-Net~\cite{ronneberger2015u} as the feature extraction module in the proposed RecPred network, but we insert three ConvLSTM~\cite{shi2015convolutional} layers to the downsampling part, the middle layers and the upsampling part, respectively. This makes the RecPred network recurrent, and therefore it is able to take advantage of temporal information in the sequential compressed frames.
After the recurrent U-Net, we use two sub-networks to generate optical flows $f_1^t$ and $f_2^t$ and the masks $m_1^t$ and $m_2^t$, which are utilized to warp and mask the previously compressed frames $\hat F_{t-1}$ and $\hat F_{t-2}$, i.e., $m_1^t\odot W_b(\hat F_{t-1}, f_1^t)$ and $m_2^t\odot W_b(\hat F_{t-2}, f_2^t)$, where $W_b$ denotes the backward warping operation and $\odot$ indicates the pixel-wise multiplication. Then, they are merged by a convolutional neural network $M$ to generate the predicted frame $\tilde F_t$. 

In addition to the hidden states that are transferred through the ConvLSTM cells, we also feed the predicted flow and masks from the previous time step as inputs into the proposed RecPred network, together with the previously compressed frames. This way, the estimated flows and masks can be expressed as 
\begin{equation}
\begin{aligned}
    &[f_1^t, f_2^t, m_1^t, m_2^t] \\
    &= \text{RecPred}(F_{t-1}, F_{t-2}, f_1^{t-1}, f_2^{t-1}, m_1^{t-1}, m_2^{t-1}, h_{t-1}),
\end{aligned}
\end{equation}
where $h_t$ is defined as the hidden states in ConvLSTM at the frame $t$. Then, the predicted frame is obtained by
\begin{equation}
\tilde F_t = M\big(m_1^t\odot W_b(\hat F_{t-1}, f_1^t),\  m_2^t\odot W_b(\hat F_{t-2}, f_2^t)\big).
\end{equation}

Next, the predicted frame $\tilde F_t$ is fed into the following steps of our ALVC framework shown in Fig.~\ref{fig:2}, i.e., the detection, compression and correction of the location error and the residual compression. Note that, in ALVC for P-frames, we use the recurrent auto-encoder and the recurrent probability model (denoted as $P$ in Fig.~\ref{fig:2})~\cite{yang2020recurrent} to compress the location error and the residual, and they work together with the proposed RecPred network to render ALVC fully recurrent in the IPPP mode. The detailed architecture of each network is shown in the \textit{Supporting Document}.

\subsection{B-frame: ALVC with bi-directional in-loop prediction}\label{B}

\begin{figure}[!t]
\centering
\includegraphics[width=\linewidth]{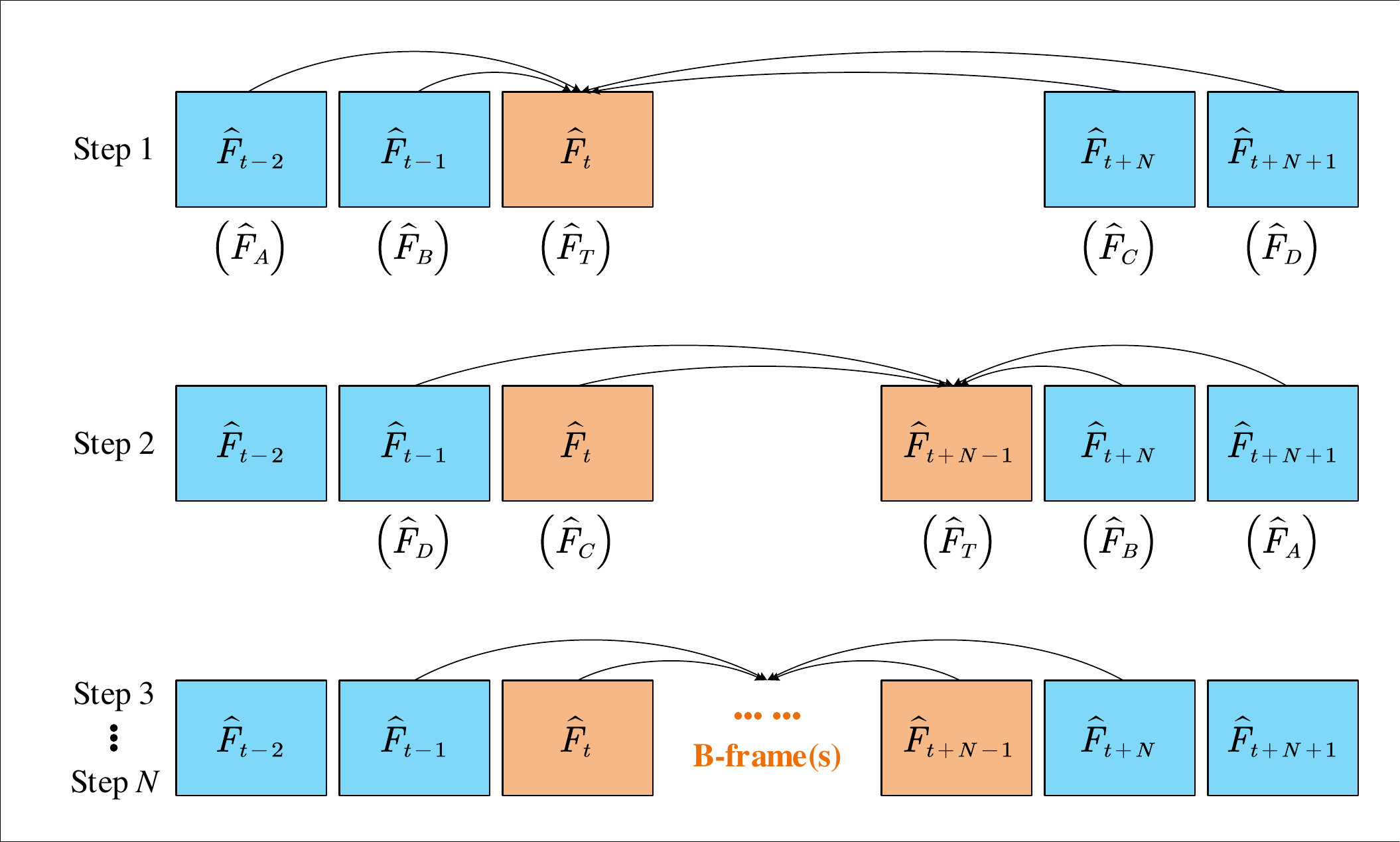}
\caption{The pipeline of ALVC for compressing consecutive B-frames (orange) between two neighboring GOPs.} \label{fig:pp}
\end{figure}

% In video compression, the B-frames are compressed using both the previous and subsequent frames as references. They are expected to have higher compression efficiency then P-frames. 
In our ALVC approach, we insert a number of consecutive B-frames between the two neighboring Group of Pictures (GOPs). B-frames are compressed using both the previous and subsequent frames as references, and therefore, they are expected to have higher compression efficiency than P-frames. We define the number of consecutive B-frames in one GOP as $N$.

\textbf{Pipeline.} Fig.~\ref{fig:pp} shows the pipeline of our ALVC approach for the compression of B-frames. As shown in Fig.~\ref{fig:pp}, $F_t$ to $F_{t+N-1}$ are the consecutive B-frames to be compressed. The previous reference frames are $\hat F_{t-2}$ and $\hat F_{t-1}$, and the subsequent reference frames are $\hat F_{t+N}$ and $\hat F_{t+N+1}$.

We start compression from $F_t$, which is the nearest frame from the previous references (Step 1). After compressing $F_t$, we use $\hat F_{t-1}$, the compressed $F_t$ (i.e., $\hat F_t$), $\hat F_{t+N}$ and $\hat F_{t+N+1}$ as references to compress the frame $F_{t+N-1}$, that is nearest from the subsequent references (Step 2). In the following, we compress $F_{t+1}$ and $F_{t+N-2}$ and so on. This pipeline is conducted recursively until all frames are compressed. 

Since all steps share the same compression network architecture, for simplification, we always define the current target frame as $F_T$ and its compressed frame as $\hat F_T$, and define the four reference frames as $\hat F_A$, $\hat F_B$, $\hat F_C$ and $\hat F_D$, among which $\hat F_A$ and $\hat F_B$ are the nearest references. As shown in Fig.~\ref{fig:pp}, in Step 1, $\hat F_A$, $\hat F_B$, $\hat F_C$ and $\hat F_D$ correspond to $\hat F_{t-2}$, $\hat F_{t-1}$, $\hat F_{t+N}$ and $\hat F_{t+N+1}$, respectively. Then, in Step~2, the target frame $F_T$ (i.e., $F_{t+N-1}$) is near the subsequent frames, so we define $\hat F_A$, $\hat F_B$, $\hat F_C$ and $\hat F_D$ in the flipped time order, i.e., as $\hat F_{t+N+1}$, $\hat F_{t+N}$, $\hat F_{t}$ and $\hat F_{t-1}$, respectively. In Step 3 to Step $N$, the frames are defined analogously.

\begin{figure*}[!t]
\centering
\includegraphics[width=.8\linewidth]{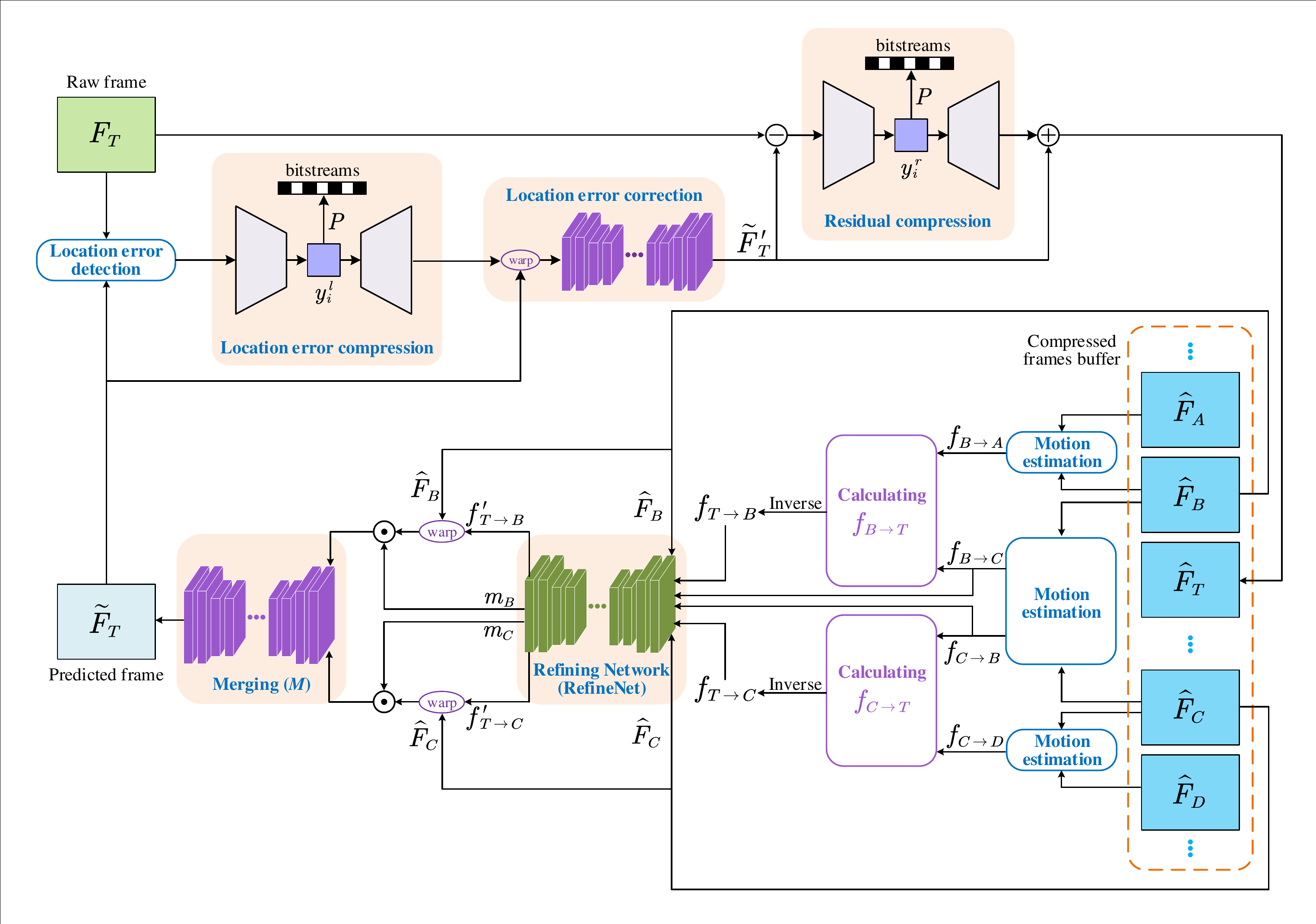}
\caption{The proposed ALVC for B-frames with bi-directional in-loop prediction.} \label{fig:bi}
\end{figure*}

\textbf{Proposed network.} Given the definitions above, Fig.~\ref{fig:bi} illustrates the ALVC network for compressing B-frame with the proposed bi-directional in-loop prediction network, which is inspired by the video interpolation work \cite{xu2019quadratic}. Specifically, we first estimate the optical flows\footnote{We denote $f_{B\to A}$ as the optical flow from $\hat F_B$ to $\hat F_A$, and other flows are denoted in the same manner.} $f_{B\to A}$, $f_{B\to C}$, $f_{C\to B}$ and $f_{C\to D}$ by the SPyNet~\cite{ranjan2017optical}, and then we use these estimated flows calculate the flows from the reference frames $\hat F_B$ and $\hat F_C$ to the target frame, i.e., $f_{B\to T}$ and $f_{C\to T}$, respectively. The details of calculating $f_{B\to T}$ and $f_{C\to T}$ are introduced in Section~\ref{C}. 

Since we use backward warping in our approach, we calculate the inverse flows of $f_{B\to T}$ and $f_{C\to T}$ as $f_{T\to B}$ and $f_{T\to C}$ to warp the reference frames $\hat F_B$ and $\hat F_C$, respectively. In what follows, a U-Net-based network is utilized as the Refining Network (RefineNet), which learns to correct the inaccuracy of the estimated flows and generate the masks ($m_B$ and $m_C$) to merge the warped reference frames, i.e.,
\begin{equation}
\begin{aligned}
    &[f'_{T\to B}, f'_{T\to C}, m_B, m_C] \\
    &= \text{RefineNet}(\hat F_B, \hat F_C, f_{T\to B}, f_{T\to C}, f_{B\to C}, f_{C\to B}),
\end{aligned}
\end{equation}
where $f'_{B\to T}$ and $f'_{B\to T}$ are the refined flows. Then, we are able to obtain 
\begin{equation}
\tilde F_T = M\big(m_B\odot W_b(\hat F_B, f'_{T\to B}),\  m_C\odot W_b(\hat F_C, f'_{T\to C})\big)
\end{equation}
as the predicted frame. Recall that $M$ indicates the CNN-based merging network, and $W_b$ denotes the backward warping operation.

The next steps for compressing $F_T$ are similar to those for P-frames, and the difference is that we replace the recurrent auto-encoder and the recurrent probability model in P-frames with the normal auto-encoder and probability model \cite{balle2017end,yang2020learning}, there is no recurrence for B-frames. The detailed architectures of each network in Fig.~\ref{fig:bi} are illustrated in the \textit{Supporting Document}. In the following section, we introduce the calculations for $f_{B\to T}$ and $f_{C\to T}$.

\subsection{Calculating $f_{B\to T}$ and $f_{C\to T}$.} \label{C}

To calculate $f_{B\to T}$ and $f_{C\to T}$ from the estimated flows $f_{B\to A}$, $f_{B\to C}$, $f_{C\to B}$ and $f_{C\to D}$, we are inspired by~\cite{xu2019quadratic} to model the temporal movement in quadratic order, i.e., 
\begin{equation}\label{qua}
f = \frac{1}{2}\cdot a\cdot \Delta t^2+v_0\cdot \Delta t,
\end{equation}
where $a$, $v_0$ and $\Delta t$ indicate the acceleration, initial velocity and time difference, respectively. Note that, since the time interval between two frames is constant in a video, we directly use the frame distance as $\Delta t$. We set the direction from $\hat F_A$ to $\hat F_D$ as the positive direction, and thus the $\Delta_t$ in this direction is positive, while that in the inverse direction is negative. 

To calculate $f_{B\to T}$, we set $\hat F_B$ as the initial point, and define the acceleration and the velocity at $\hat F_B$ as $a_B$ and $v_B$, respectively. Note that, according to the definition in Fig.~\ref{fig:pp}, we always have $\Delta t_{B\to A} = -1$, $\Delta t_{B\to T} = 1$ and $\Delta t_{C\to D} = 1$. Thus, given~\eqref{qua}, we can express $f_{B\to A}$ and $f_{B\to C}$ as
\begin{equation}\label{qua1}
\begin{aligned}
    f_{B\to A} &= \frac{1}{2}\cdot a_B\cdot (-1)^2+v_B\cdot (-1), \\
    f_{B\to C} &= \frac{1}{2}\cdot a_B\cdot \Delta t_{B\to C}^2+v_B\cdot \Delta t_{B\to C},
\end{aligned}
\end{equation}
and hence, we can obtain $a_B$ and $v_B$ as
\begin{equation}\label{av1}
\begin{aligned}
a_B &= \frac{2\cdot(\Delta t_{B\to C} \cdot f_{B\to A} + f_{B\to C})}{\Delta t_{B\to C}^2 + \Delta t_{B\to C}}, \\
v_B &= \frac{-\Delta t_{B\to C}^2\cdot f_{B\to A} + f_{B\to C}}{\Delta t_{B\to C}^2 + \Delta t_{B\to C}}.
\end{aligned}
\end{equation}
Finally, since $\Delta t_{B\to T} = 1$ is known, $f_{B\to T}$ can be calculated as 
\begin{equation}
    f_{B\to T} = \frac{1}{2}\cdot a_B\cdot 1^2+v_B\cdot 1,
\end{equation}
with $a_B$ and $v_B$ shown in \eqref{av1}.

Similarly, for calculating $f_{C\to T}$ from $f_{C\to B}$ and $f_{C\to D}$, we set $\hat F_C$ as the initial point, and define the acceleration and the velocity as $a_C$ and $v_C$, respectively. Recall $\Delta t_{C\to D} = 1$, and thus $f_{C\to B}$ and $f_{C\to D}$ can be modelled as
\begin{equation}\label{qua2}
\begin{aligned}
    f_{C\to B} &= \frac{1}{2}\cdot a_C\cdot \Delta t_{C\to B}^2+v_c\cdot \Delta t_{C\to B},\\
    f_{C\to D} &= \frac{1}{2}\cdot a_C \cdot 1^2 +v_C \cdot 1, 
\end{aligned}
\end{equation}
Then, $a_C$ and $v_C$ are obtained as
\begin{equation}\label{av2}
\begin{aligned}
a_C &= \frac{2\cdot(-\Delta t_{C\to B} \cdot f_{C\to D} + f_{C\to B})}{\Delta t_{C\to B}^2 - \Delta t_{C\to B}}, \\
v_C &= \frac{\Delta t_{C\to B}^2\cdot f_{C\to D} - f_{C\to B}}{\Delta t_{C\to B}^2 - \Delta t_{C\to B}}.
\end{aligned}
\end{equation}
Finally, $f_{C\to T}$ is calculated as
\begin{equation}\label{result2}
    f_{C\to T} = \frac{1}{2}\cdot a_C\cdot \Delta t_{C\to T}^ 2+v_C\cdot \Delta t_{C\to T}.
\end{equation}
Note that, in \eqref{qua2}, \eqref{av2} and \eqref{result2}, $\Delta t_{C\to B}$ and $\Delta t_{C\to T}$ are opposite to the positive direction, so they are negative values.

Recall that we defined $N$ as the number of consecutive B-frames in a GOP, and those B-frames are compressed by Step 1 to Step $N$ in accordance with the pipeline in Fig.~\ref{fig:pp}. As such, in a specific step, e.g., Step $n$, we have $\Delta t_{B\to C} = N - n + 2$, $\Delta t_{C\to T} = -(N - n + 1)$ and $\Delta t_{C\to B} = -(N -n + 2)$. Put these into the equations \eqref{qua1} to \eqref{result2}, the optical flows $f_{B\to T}$ and $f_{C\to T}$ can be calculated. Then, as described in Section~\ref{B}, they are fed into the following modules to generate the predicted frame $\tilde F_T$.

\subsection{Training}
We use Vimeo-90k dataset~\cite{xue2019video} as the training set. We first pre-train the in-loop prediction networks on raw frames. Afterwards, we jointly train the whole ALVC network in an end-to-end manner with the rate-distortion loss function
\begin{equation}\label{loss}
    \mathcal{L} = \sum_t R(y^l_t) + R(y^r_t) + \lambda \cdot  D(\hat F_t, F_t),
\end{equation}
where $R(y^l_t)$ and $R(y^r_t)$ are the estimated bit-rates of the latent representations for compressing the location error and residual, respectively. $D$ indicates the distortion between compressed and raw frames and $\lambda$ is the hyper-parameter to control the rate-distortion trade-off. In this paper, we use the Mean Squared Error (MSE) and the Multi-scale Structural SIMilarity (MS-SSIM) index as $D$ to train the PSNR and MS-SSIM models, respectively. For the PSNR models, we set $\lambda$ to 256, 512, 1024 and 2048, and for the MS-SSIM models, $\lambda$ is set as 8, 16, 32 and 64.

\section{Experiments}

\subsection{Settings}

We follow the previous learned video compression approaches~\cite{wu2018video,lu2019dvc,cheng2019learning,habibian2019video,yang2020learning,agustsson2020scale,golinski2020feedback,yang2020recurrent,hu2021fvc,li2021deep} to evaluate the performance on the JCT-VC~\cite{bossen2013common}, UVG~\cite{UVG} and VTL~\cite{seeling2014video} datasets. The Classes B, C and D in JCT-VC contain normal videos with resolution of $1920\times 1080$, $832\times 480$ and $416\times 240$, respectively. JCT-VC Class E contains conversational videos with the resolution of $1280\times 720$. 
The UVG dataset has the videos at $1920\times 1080$. For the VTL dataset, we follow~\cite{lu2020content,hu2020improving,yang2020recurrent,hu2021fvc} to test on the first 300 frames of the videos in CIF format ($352\times 288$) for fair comparisons. In our approach, we set the GOP size as 13 frames, with one I-frame (compressed by VTM~\cite{VTM} in the PSNR model and by \cite{cheng2020learned} in the MS-SSIM model), ten P-frames and two B-frames ($N=2$). In the ablation studies, we also analyse the performance of various GOP sizes, the number of consecutive B-frames, and the impact of the I-frame codec. 

{In our experiments, we compare the proposed ALVC method with the existing learned video compression methods, including DVC~\cite{lu2019dvc}, HLVC~\cite{yang2020learning}, Agustsson~et al.~\cite{agustsson2020scale}, RLVC~\cite{yang2020recurrent}, Lu~et al.~\cite{lu2020content}, Hu~et al.~\cite{hu2020improving}, FVC~\cite{hu2021fvc}, Liu~et al.~\cite{liu2022end} and DCVC~\cite{li2021deep}. Then, we also compare ALVC with various configurations of x265. Specifically, the PSNR model of ALVC is compared with x265 (LDP) and x265 (B-frames). Their detailed settings are shown as follows:
\begin{itemize}
\item x265 (LDP) \\ \texttt{ffmpeg -pix\_fmt yuv420p -s HxW -i input.yuv -r FR -c:v libx265\\-tune zerolatency
-x265-params  "crf=CRF:keyint=13" output.mkv}
\item x265 (B-frame) \\\texttt{ffmpeg -pix\_fmt yuv420p -s HxW -i input.yuv -r FR -c:v libx265 \\
-x265-params "b-adapt=0:bframes=2:\\b-pyramid=1:crf=CRF:keyint=13" output.mkv}
\end{itemize}
In these settings, \texttt{H}, \texttt{W} and \texttt{FR} refer to height, width and frame rate, respectively. The quality factor \texttt{CRF} ranges from 15 to 27. In x265 (B-frame), we use the same GOP size (=13) and the same number of B-frames ($N=2$) as ALVC. ``\texttt{b-adapt=0}'' indicates a fixed GOP, and ``\texttt{b-pyramid=1}'' means that B-frames can serve as reference frames. These are consistent with our ALVC approach.}

\begin{table*}[!t]
\scriptsize
\centering
\caption{BDBR ($\%$) calculated by PSNR with the anchor of \textit{x265 (B-frame)}. \textbf{Bold} indicates the best results in \emph{learned} approaches.
}\label{tab:bdbr_psnr}
\setlength{\tabcolsep}{2pt}

\begin{tabular}{c@{\hskip 2ex}rrrrrrr@{\hskip 0ex}rrrrrr}
\cmidrule[\heavyrulewidth]{1-14}  % 
& \multicolumn{7}{c}{Learned}&\multicolumn{6}{c}{Non-learned} \\
\cmidrule(r{.5em}){2-8}
\cmidrule(l{.5em}){9-14} 

& \multicolumn{1}{c}{DVC~\cite{lu2019dvc}} &
\multicolumn{1}{c}{HLVC~\cite{yang2020learning}}&
\multicolumn{1}{c}{RLVC~\cite{yang2020recurrent}} &
\multicolumn{1}{c}{Liu~et al.~\cite{liu2022end}} &
\multicolumn{1}{c}{FVC~\cite{hu2021fvc}} &
\multicolumn{1}{c}{DCVC~\cite{li2021deep}} &
\multicolumn{1}{c}{ALVC} &
\multicolumn{1}{c}{x265} & 
\multicolumn{1}{c}{x265} &
\multicolumn{1}{c}{HM 16.20} &
\multicolumn{1}{c}{HM 16.20} &
\multicolumn{1}{c}{VTM 14.1} &
\multicolumn{1}{c}{VTM 14.1} \\
\multicolumn{1}{c}{Dataset} & \multicolumn{1}{c}{(CVPR'19)}&  
\multicolumn{1}{c}{(CVPR'20)}&
\multicolumn{1}{c}{(JSTSP'21)}&
\multicolumn{1}{c}{(TCSVT'22)}&
\multicolumn{1}{c}{(CVPR'21)}&
\multicolumn{1}{c}{(NIPS'21)}&
\multicolumn{1}{c}{\ \ (ours)\ \ \ }&
\multicolumn{1}{c}{(LDP)}&
\multicolumn{1}{c}{(B-frame)}&
\multicolumn{1}{c}{(same GOP)} &
\multicolumn{1}{c}{(default)} &
\multicolumn{1}{c}{(same GOP)} &
\multicolumn{1}{c}{(default)}\\

\cmidrule{1-14} 
    {Class B}  & $47.73$&$28.71$&$9.94$&$8.01$&$12.79$&$\mathbf{-14.60}$&$-10.86\ \ \ \ \  $&$19.81$&$0.00$&$-10.53$&$-30.30$&$-39.59$&$-53.65$ 
 \\

\cmidrule{1-14} 
    {Class C}  & $58.77$&$37.48$&$20.53$&$36.43$&$8.68$&$2.94$&$\mathbf{-0.21}\ \ \ \ \  $&$15.13$&$0.00$&$-7.54$&$-30.35$&$-30.99$&$-49.96$
 \\

\cmidrule{1-14} 
    {Class D} & $49.56$ &	$13.41$ 	&$-6.14$&$59.93$&$6.57$&$-5.29$&$\mathbf{-24.03}\ \ \ \ \ $&$18.33$&$0.00$&$-8.41$&$-32.13$&$-28.97$&$-48.01$
 \\
\cmidrule{1-14}

    {UVG}  & $44.89$&$31.74$&$13.80$&$3.54$&$10.09$&$\mathbf{-8.60}$&$-3.29\ \ \ \ \  $&$15.08$&$0.00$&$-16.12$&$-38.80$&$-42.57$ & $-59.03$
 \\
    
\cmidrule[\heavyrulewidth]{1-14}
    \textbf{Average} & $50.24$&$27.83$&$9.53$&$26.98$&$9.53$&$-6.39$&$\mathbf{-9.60}\ \ \ \ \ $&$17.09$&$0.00$&$-10.65$&$-32.89$&$-35.53$ &$-52.66$
 \\
    
\cmidrule[\heavyrulewidth]{1-14} 

\end{tabular}
\end{table*}

\begin{figure*}[!t]
\justify
\subfigure[The PSNR performance compared with learned approaches.]{\includegraphics[width=.48\linewidth]{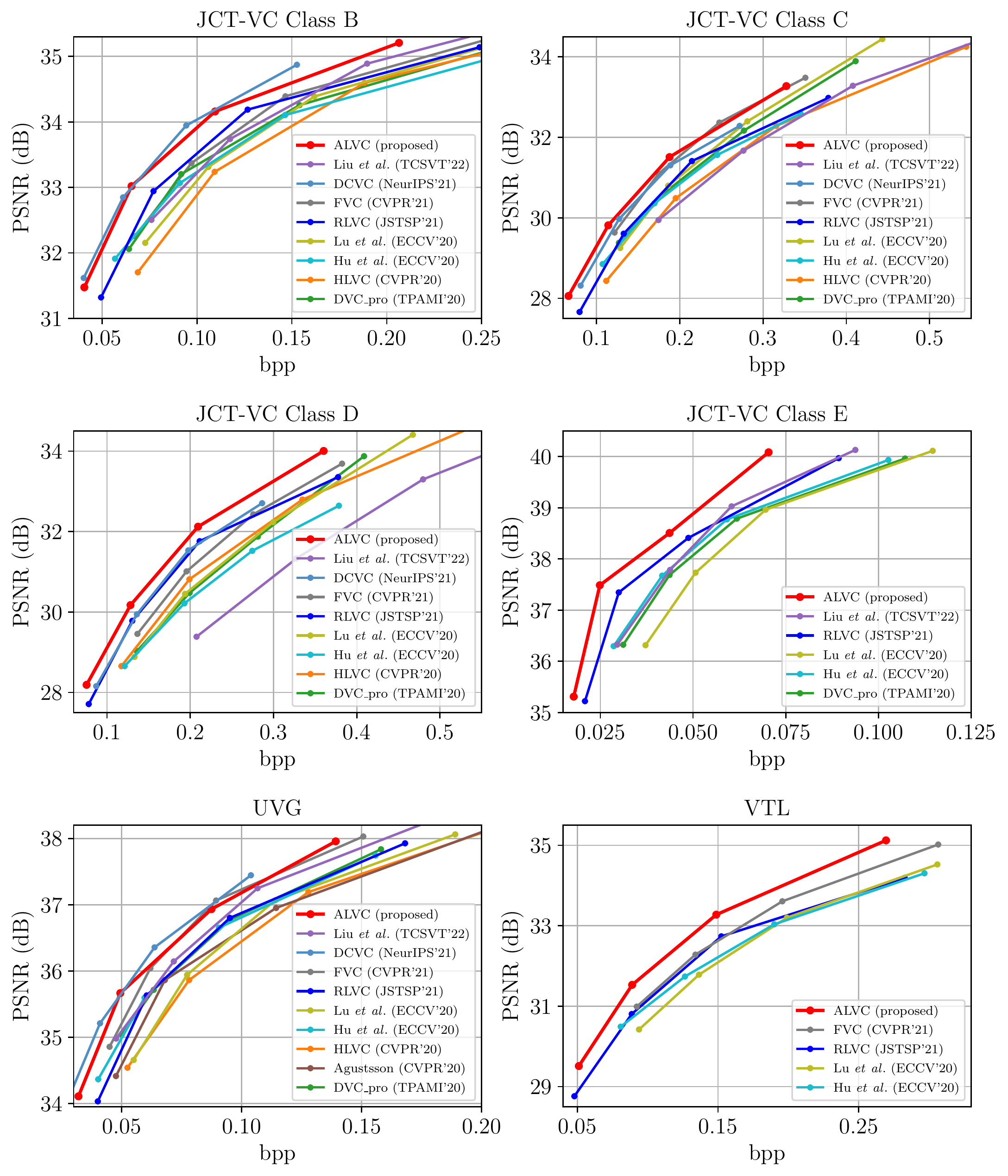}}
\subfigure[The PSNR performance compared with handcrafted algorithms.]{\includegraphics[width=.48\linewidth]{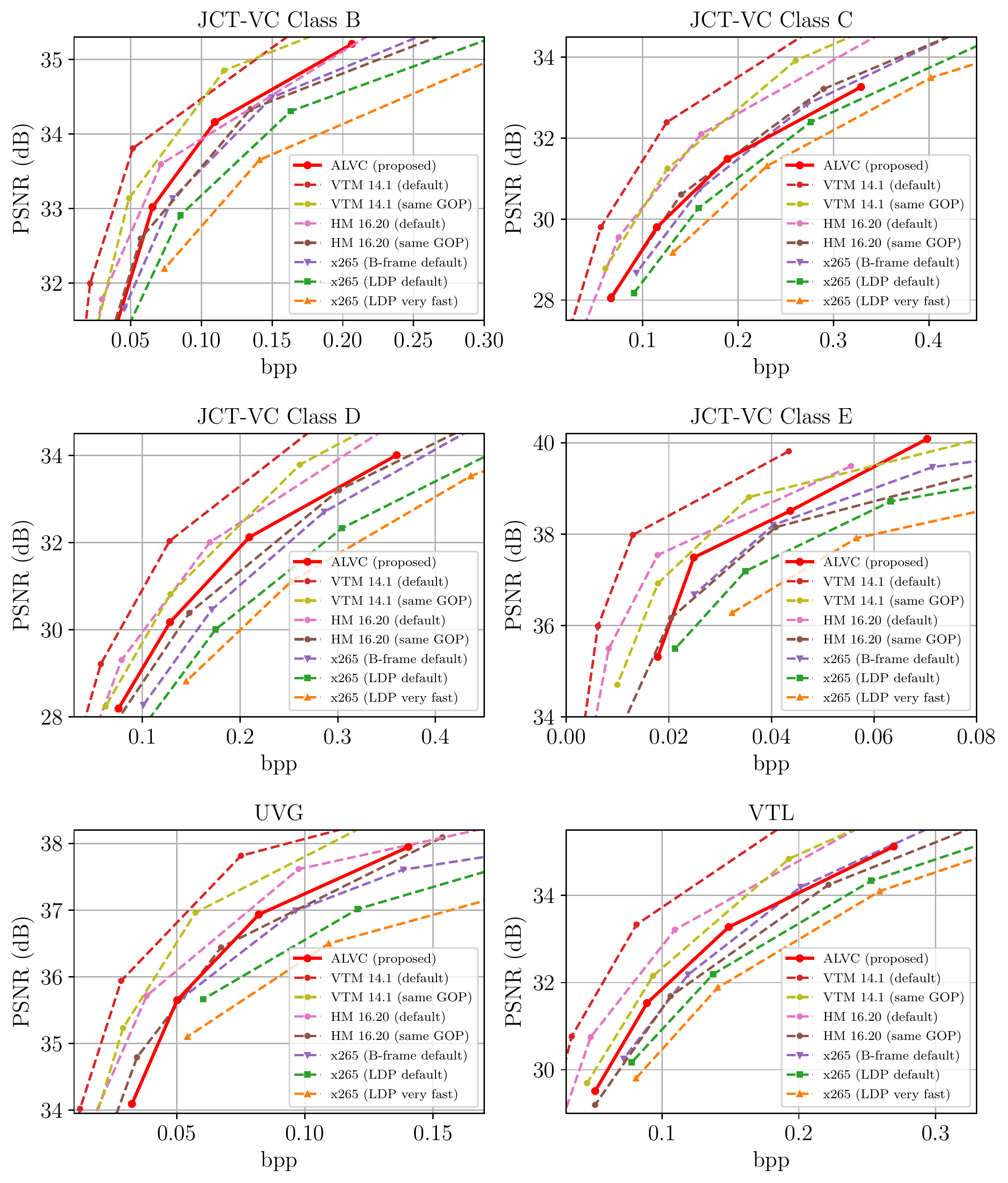}}
\caption{The PSNR performance compared with previous learned and handcrafted algorithms.} \label{fig:rd}
\end{figure*}

{In the comparison with the MS-SSIM model of ALVC, we further include the SSIM-tuned x265 with the default and the slowest modes. The detailed settings are shown as follows:
\begin{itemize}
% \item x265 (LDP default) \\ \texttt{ffmpeg -pix\_fmt yuv420p -s HxW -i input.yuv -r FR -c:v libx265\\-tune zerolatency
% -x265-params  "crf=CRF" output.mkv}
% \item x265 (default) \\ \texttt{ffmpeg -pix\_fmt yuv420p -s HxW -i input.yuv -r FR -c:v libx265\\
% -x265-params  "crf=CRF" output.mkv}
\item x265 (SSIM default) \\ \texttt{ffmpeg -pix\_fmt yuv420p -s HxW -i input.yuv -r FR -c:v libx265\\
-tune ssim -x265-params  "crf=CRF" output.mkv}
% \item x265 (slowest) \\ \texttt{ffmpeg -pix\_fmt yuv420p -s HxW -i input.yuv -r FR -c:v libx265\\
% -preset placebo -x265-params  "crf=CRF" output.mkv}
\item x265 (SSIM slowest) \\ \texttt{ffmpeg -pix\_fmt yuv420p -s HxW -i input.yuv -r FR -c:v libx265\\
-preset placebo -tune ssim -x265-params  "crf=CRF" output.mkv}
\end{itemize}
In these settings, ``\texttt{-preset placebo}'' is the slowest mode (the best performance) of x265, and ``\texttt{-tune ssim}'' indicates the SSIM-tuned x265. As far as we know, x265 (SSIM slowest) has the \emph{best} MS-SSIM performance that x265 may reach. Besides, we also compare with x265 (default) and x265 (slowest), in which ``\texttt{-tune ssim}'' is removed from x265 (SSIM default) and x265 (SSIM slowest), respectively.}

{Moreover, we also compare ALVC with the HEVC test model (HM 16.20) and the VVC test model (VTM 14.1). First, the HM 16.20 and VVC 14.1 models are tested under the same GOP structure as ALVC, i.e., the same number of I-, P-, and B-frames in each GOP and the P- and B-frames are with flat quality. These settings are denoted as HM 16.20 (same GOP) and VVC 14.1 (same GOP), respectively. Then, we also compare with the default settings of HM and VTM, which are defined as HM 16.20 (default) and VTM 14.1 (default).}

\begin{figure}[!t]
\centering
\includegraphics[width=1\linewidth]{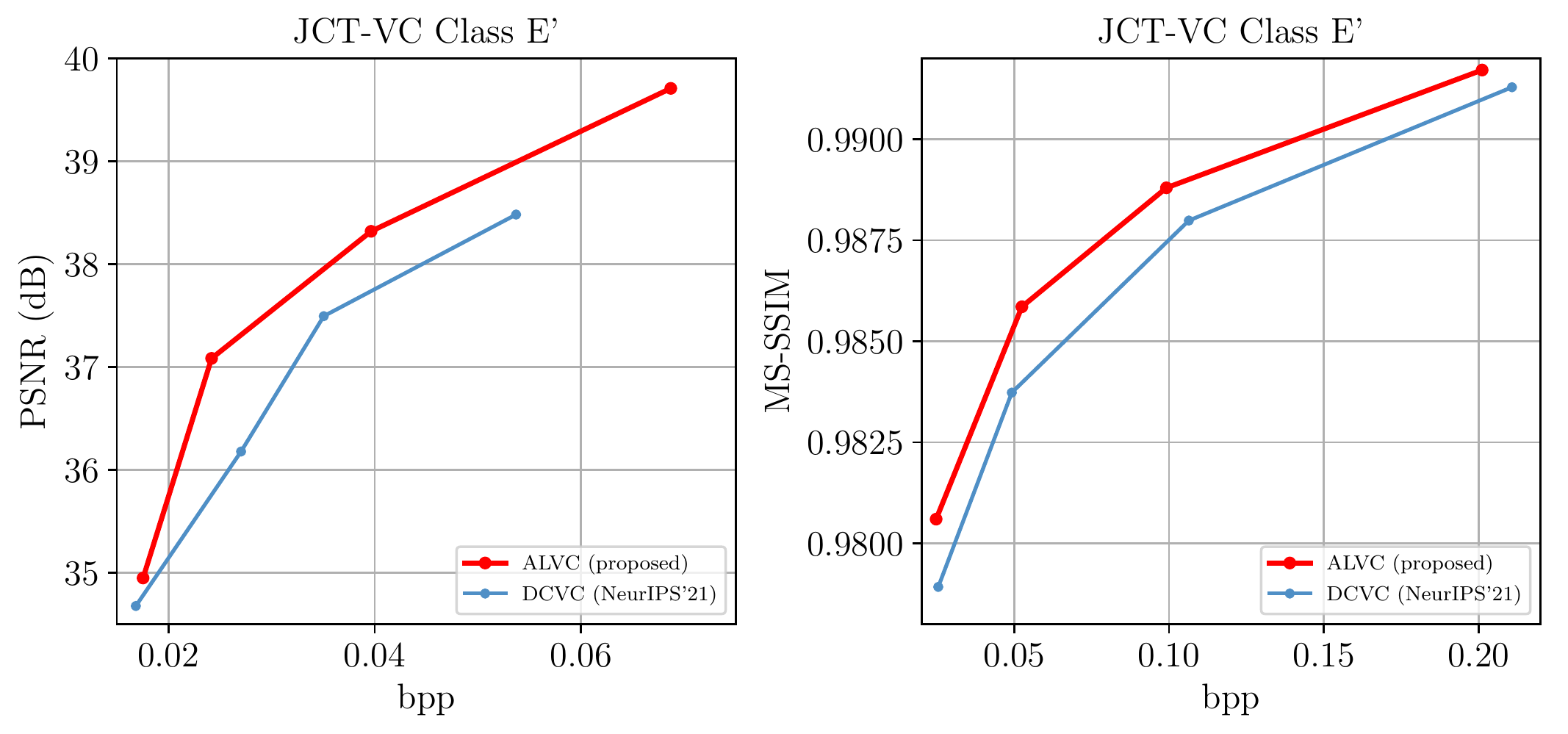}
\caption{The rate-distortion performance compared with DCVC~\cite{li2021deep} on the JCT-VC Class E' (\textit{FourPeople}, \textit{Johnny} and \textit{KristenAndSara}).} \label{fig:rd_e}
\end{figure}

\begin{table*}[!t]
\scriptsize
\centering
\caption{BDBR ($\%$) calculated by MS-SSIM with the anchor of \textit{x265 (SSIM slowest)}. \textbf{Bold} indicates the best results in \emph{learned} approaches.
}\label{tab:bdbr2}
\setlength{\tabcolsep}{2pt}

\begin{tabular}{c@{\hskip 2ex}rrrrrrr@{\hskip 0ex}rcrrrr}
\cmidrule[\heavyrulewidth]{1-14}  % 
& \multicolumn{7}{c}{Learned}&\multicolumn{6}{c}{Non-learned} \\
\cmidrule(r{.5em}){2-8}
\cmidrule(l{.5em}){9-14} 

& \multicolumn{1}{c}{DVC~[5]} &
\multicolumn{1}{c}{HLVC~[7]}&
\multicolumn{1}{c}{RLVC~[9]} &
\multicolumn{1}{c}{Liu~et al.~[43]} &
\multicolumn{1}{c}{FVC~[10]} &
\multicolumn{1}{c}{DCVC~[44]} &
\multicolumn{1}{c}{ALVC} &
% \multicolumn{1}{c}{x265} & 
\multicolumn{1}{c}{x265} &
\multicolumn{1}{c}{x265} &
\multicolumn{1}{c}{HM 16.20} &
\multicolumn{1}{c}{HM 16.20} &
\multicolumn{1}{c}{VTM 14.1} &
\multicolumn{1}{c}{VTM 14.1} \\
\multicolumn{1}{c}{Dataset} & 
\multicolumn{1}{c}{(CVPR'19)}&  
\multicolumn{1}{c}{(CVPR'20)}&
\multicolumn{1}{c}{(JSTSP'21)}&
\multicolumn{1}{c}{(TCSVT'22)}&
\multicolumn{1}{c}{(CVPR'21)}&
\multicolumn{1}{c}{(NIPS'22)}&
\multicolumn{1}{c}{\ \ (ours)\ \ \ }&
% \multicolumn{1}{c}{default}&
\multicolumn{1}{c}{(SSIM)}&
\multicolumn{1}{c}{(SSIM slowest)}&
\multicolumn{1}{c}{(same GOP)} &
\multicolumn{1}{c}{(default)} &
\multicolumn{1}{c}{(same GOP)} &
\multicolumn{1}{c}{(default)}\\

\cmidrule{1-14} 
    {Class B}  & $94.31$&$26.55$&$2.09$&$4.17$&$-10.15$&$-3.24$&$\mathbf{-13.04}\ \ \ \ \ $&$20.11$&$0.00$&$70.76$&$32.07$&$10.04$&$-14.61$
\\
\cmidrule{1-14}
    {Class C}  & $74.45$&$42.87$&$18.33$&$11.62$&$0.77$&$7.02$&$\mathbf{-0.95}\ \ \ \ \ $&$20.72$&$0.00$&$63.89$&$25.41$&$20.97$&$-10.92$\\
\cmidrule{1-14}
    {Class D}  & $56.20$&$-11.94$&$-1.16$&$5.45$&$-6.67$&$-2.78$&$\mathbf{-18.77}\ \ \ \ \ $&$19.83$&$0.00$&$88.23$&$31.22$&$36.68$&$-5.33 
$\\

\cmidrule{1-14}

    {UVG}  &  $184.94$&$16.84$&$-2.81$&$-9.52$&$-13.88$&$\mathbf{-16.54}$&$-15.44\ \ \ \ \ $&$39.94$&$0.00$&$133.00$&$28.47 
$& $55.40$ & $-16.31$\\

\cmidrule[\heavyrulewidth]{1-14}
    \textbf{Average} &  $102.47$&$18.58$&$4.11$&$2.93$&$-7.48$&$-3.89$&$\mathbf{-12.05}\ \ \ \ \ $&$25.15$&$0.00$&$88.97$&$29.29 
$& $30.77$ & $-11.79$\\
\cmidrule[\heavyrulewidth]{1-14} 

\end{tabular}
\end{table*}

\begin{figure*}[!t]
\justify
\subfigure[The MS-SSIM performance compared with learned approaches.]{\includegraphics[width=.48\linewidth]{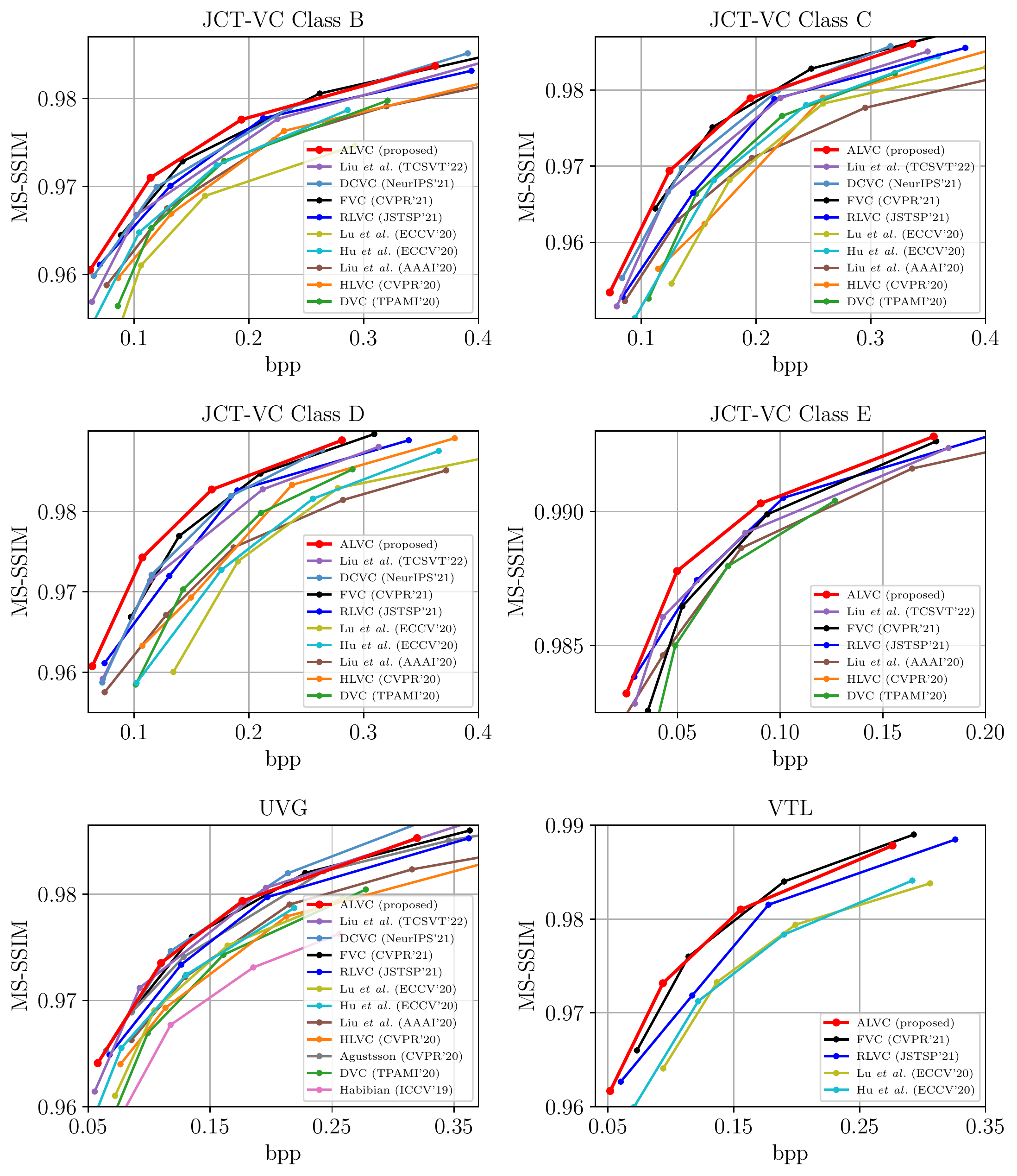}}
\subfigure[The MS-SSIM performance compared with handcrafted algorithms.]{\includegraphics[width=.48\linewidth]{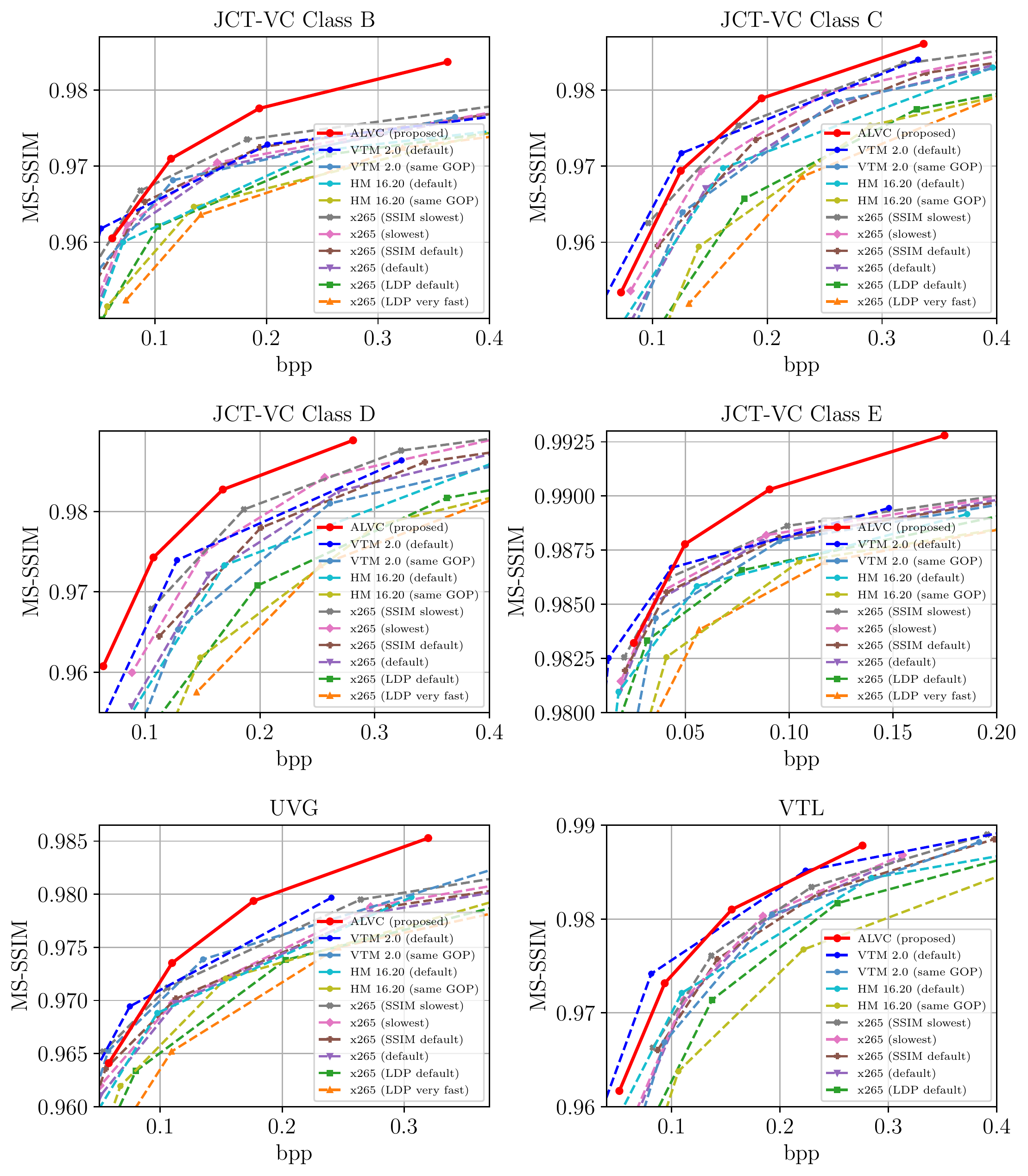}}
\caption{The MS-SSIM performance compared with previous learned and handcrafted algorithms.} \label{fig:ssim}
\end{figure*}

\subsection{Rate-distortion performance in terms of PSNR}

{Table~\ref{tab:bdbr_psnr} shows the Bj{\o}ntegaard Delta Bit-Rate (BDBR)~\cite{bjontegaard} performance\footnote{Since some works do not provide the detailed data on each video of Class E and VTL, we compare BDBR on UVG and JCT-VC Classes B, C and D.} (lower is better) calculated by PSNR with the anchor of x265 (B-frames). It can be seen from Table~\ref{tab:bdbr_psnr} that our ALVC approach outperforms DVC~\cite{lu2019dvc}, HLVC~\cite{yang2020learning}, RLVC~\cite{yang2020recurrent}, Liu~et al.~\cite{liu2022end}, FVC~\cite{hu2021fvc}, x265 (LDP) and x265 (B-frame) on \textit{all} datasets. We also beat DCVC~\cite{li2021deep} on Class C and Class D, and the average BDBR of our ALVC approach is $3.21\%$ better than DCVC~\cite{li2021deep}. Similar results can also be observed on the rate-distortion curves illustrated in Fig.~\ref{fig:rd}-(a). Besides, we can also see from Fig.~\ref{fig:rd} that we perform best among learned approaches on Class E and VTL. Note that we follow most previous learned video compression works~\cite{lu2019dvc,yang2020learning,yang2020recurrent,liu2022end,hu2021fvc} to define the sequences \textit{Vidyo1}, \textit{Vidyo3} and \textit{Vidyo4} as Class E, while DCVC~\cite{li2021deep} uses the sequences \textit{FourPeople}, \textit{Johnny} and \textit{KristenAndSara} as Class E. In this paper, we define \textit{FourPeople}, \textit{Johnny} and \textit{KristenAndSara} as Class E' and Fig.~\ref{fig:rd_e} shows that our ALVC obviously outperforms DCVC on Class E'.}

{When comparing ALVC with HM 16.20 (same GOP), we perform better on Class B and Class D, but averagely worse than HM 16.20 (same GOP) by around $1\%$ of BDBR. Meanwhile, Fig.~\ref{fig:rd}-(b) indicates that we also outperform HM 16.20 (same GOP) on Class E and VTL. However, Table~\ref{tab:bdbr_psnr} and Fig.~\ref{fig:rd}-(b) show that our ALVC is not able to catch up with the performance of HM 16.20 (default), VTM 14.1 (same GOP) and VTM 14.1 (default), although we achieve the state-of-the-art performance in learned approaches.}

\subsection{Rate-distortion performance in terms of MS-SSIM}

{The BDBR performance in terms of MS-SSIM is shown in Table~\ref{tab:bdbr2}, which uses the performance of x265 (SSIM slowest) as the anchor. It can be seen from this table that our ALVC approach performs better than DVC~\cite{lu2019dvc}, HLVC~\cite{yang2020learning}, RLVC~\cite{yang2020recurrent}, Liu~et al.~\cite{liu2022end}, FVC~\cite{hu2021fvc}, x265 (SSIM), x265 (SSIM slowest), HM 16.20 (same GOP), HM 16.20 (default) and VTM 14.1 (same GOP) on \textit{all} datasets. We also outperform DCVC~\cite{li2021deep} on Classes B, C and D. We are slight worse than DCVC on UVG, but we averagely outperform DCVC by more than $8\%$ on BDBR. Similar results can be observed from the rate-distortion curves in Fig.~\ref{fig:ssim}-(a). Fig.~\ref{fig:rd_e} also shows that our MS-SSIM performance is also obviously better than DCVC on Class E'. Besides, we are better than VTM 14.1 (default) on Class D in terms of MS-SSIM, but fail to beat VTM 14.1 (default) on other datasets. However, the average MS-SSIM performance of our ALVC approach ($-12.05\%$) is slightly better than VTM 14.1 (default) ($-11.79\%$). Fig.~\ref{fig:ssim}-(b) shows that we significantly outperform the handcrafted codecs at high bit-rates on MS-SSIM, and we are even better than VTM 14.1 (default) on all datasets at high bit-rates. 
In conclusion, we reach the state-of-the-art performance on MS-SSIM among learned approaches, and our MS-SSIM performance is comparable with and slightly better than VTM 14.1 (default) and we also beat all other settings of handcrafted codecs.}

\subsection{Visual results}

\begin{figure*}[!t]
\centering
\includegraphics[width=\linewidth]{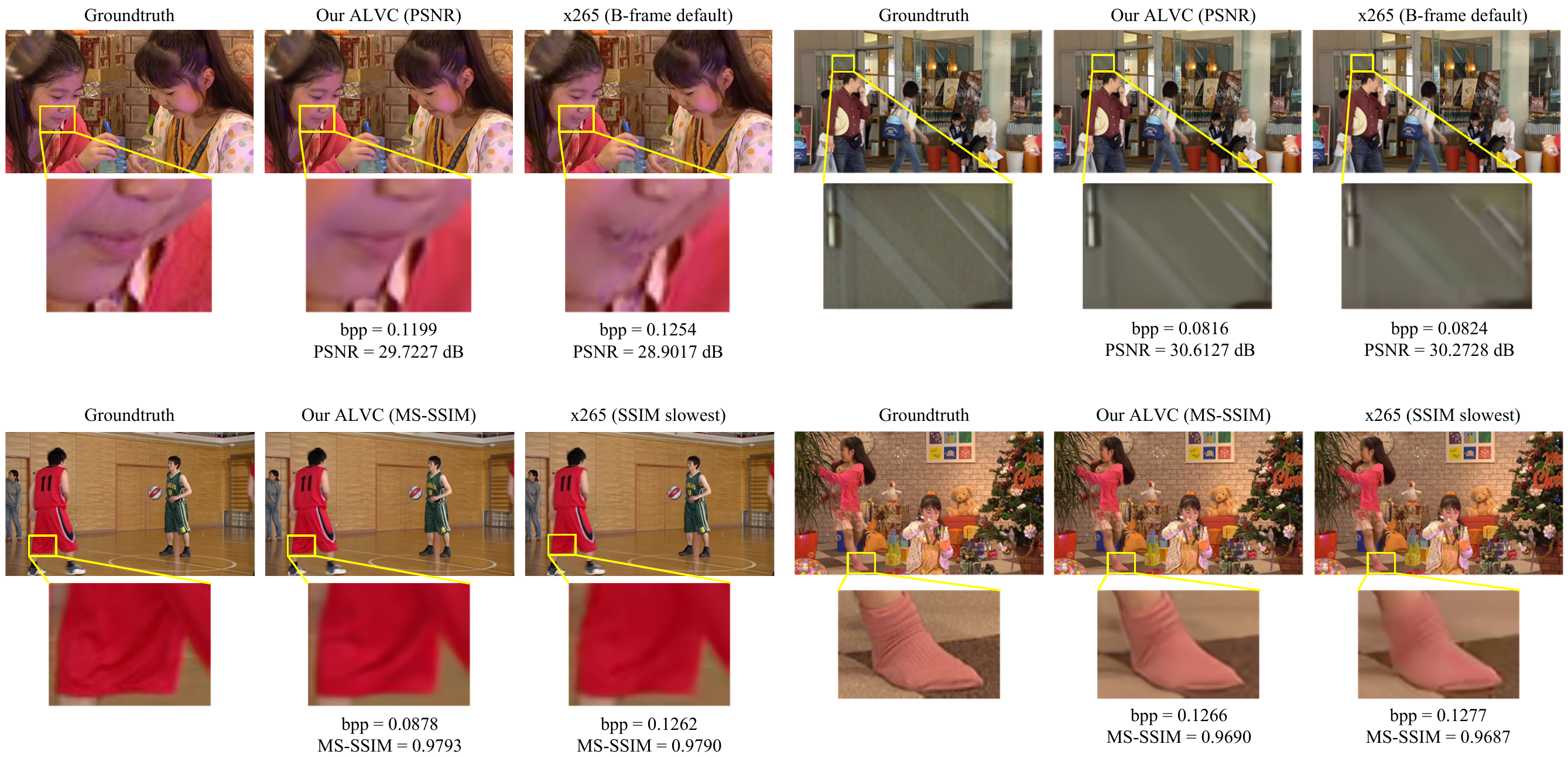}
\caption{Visual results of the PSNR model (first raw) and the MS-SSIM model (second raw) of the proposed ALVC in comparison with x265.} \label{fig:vis}
\end{figure*}

We show the visual results of our ALVC approach in Fig.~\ref{fig:vis}. We compare the results of our PSNR-optimized model with x265 (B-frames default), and compare our MS-SSIM-optimized model with x265 (SSIM slowest). It can be seen in Fig.~\ref{fig:vis} that our PSNR-optimized ALVC model obviously achieves fewer compression artifacts than x265 (B-frames default), and our MS-SSIM-optimized model maintains more textual details than x265 (SSIM slowest). 

\subsection{Time complexity}

\begin{table}[!t]
\centering
\caption{Time complexity (second per frame) on 1080p videos
}\label{tab:time}

\begin{tabular}{cccccc}
\cmidrule[\heavyrulewidth]{1-6}  % 
& DVC &
RLVC &
FVC &
DCVC &
\multicolumn{1}{c}{ALVC} \\
 & \cite{lu2019dvc}  & \cite{yang2020recurrent} & \cite{hu2021fvc}& \cite{li2021deep}&
\multicolumn{1}{c}{(proposed)} \\
\cmidrule{1-6} 
    Encoding& $0.558$ & $0.812$ & $0.548${*} & $0.826$ & $1.151$\\
\cmidrule{1-6}
    Decoding& $0.278$ & $0.374$ & $-^\dagger$ & $46.2$ & $0.736$\\
\cmidrule[\heavyrulewidth]{1-6} 
\multicolumn{6}{l}{{*} No open-sourced codes, so we copy the values from~\cite{hu2021fvc}}\\  \multicolumn{6}{l}{$\dagger$ No open-sourced codes, and also not reported}\\
\end{tabular}
\end{table}

We evaluate the time complexity of our ALVC approach on an NVIDIA TITAN Xp GPU. The average encoding time and decoding time on 1080p videos are listed in Table~\ref{tab:time}. {The FVC is not open-sourced, so we are not able to evaluate the speed of FVC under the same hardware as our approach. In Table~\ref{tab:time}, the encoding time of FVC is directly copied from the FVC paper~\cite{hu2021fvc}, which are tested on an NVIDIA GeForce 2080 Ti GPU according to \cite{hu2021fvc}.  Besides, FVC~\cite{hu2021fvc} does not report the decoding time.} The encoding and decoding time of DVC, RLVC and DCVC are tested on the same GPU as our ALVC approach, using the open-sourced codes of these methods. 

It can be seen from Table~\ref{tab:time} that the encoding time of ALVC is 1.151 seconds per frame, and our decoding is 0.736 seconds per frame. The time complexity of our approach is higher than FVC and RLVC. {Our encoding speed is also slower than DCVC, but our decoding speed is significantly faster than DCVC, since DCVC uses a spatial auto-regressive entropy model, which leads to slow decoding.}
Recall that our ALVC approach achieves better rate-distortion performance. Table~\ref{tab:bdbr_psnr} shows that our average BDBR is $>3\%$ better than DCVC and $>19\%$ better than FVC and RLVC in terms of PSNR. Table~\ref{tab:bdbr2} shows that our BDBR is $>8\%$ better than DCVC, $>4.5\%$ better than FVC and $>16\%$ better than RLVC in terms of MS-SSIM.

{In the handcrafted video coding algorithms, x265 is optimized towards speed, and therefore the x265 (B-frame) has an encoding time of 0.05 seconds per frame for the 1080p videos on an Intel(R) Core(TM) i7-8700 CPU. For HM 16.20 and VTM 14.1, the encoding speed is slow. The encoding time of HM 16.20 is 23.0 s, 17.3 s, 14.4 s and 12.7 s per 1080p frame at QP = 22, 27, 32 and 37, respectively. VTM 14.1 has a slower speed, whose encoding time is 360.8 s, 185.4 s, 105.9 s and 62.3 s per 1080p frame at QP = 22, 27, 32 and 37, respectively.}

\begin{figure*}[!t]
\centering
\subfigure[Visualizing the location error and temporal motion. {We use the same color wheel as \cite{baker2011database}.} ]{\includegraphics[width=.65\linewidth]{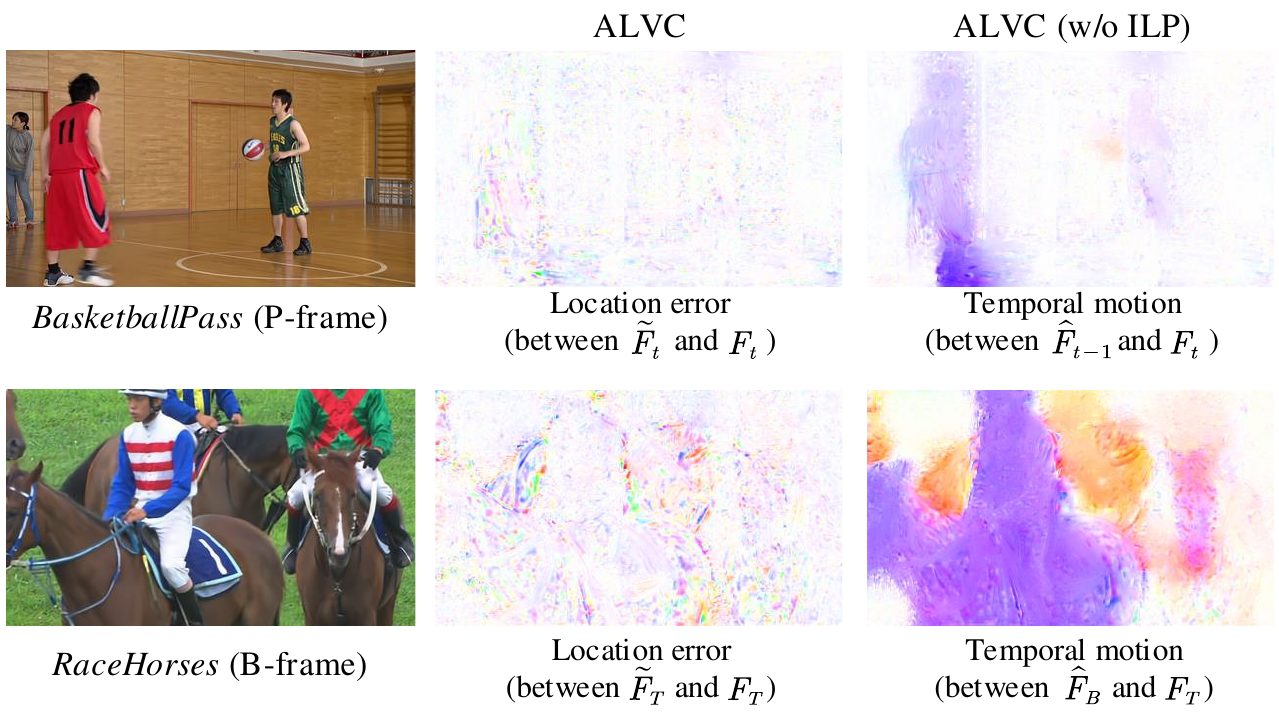}}
\subfigure[Ablation performance on frame prediction ]{\includegraphics[width=.34\linewidth]{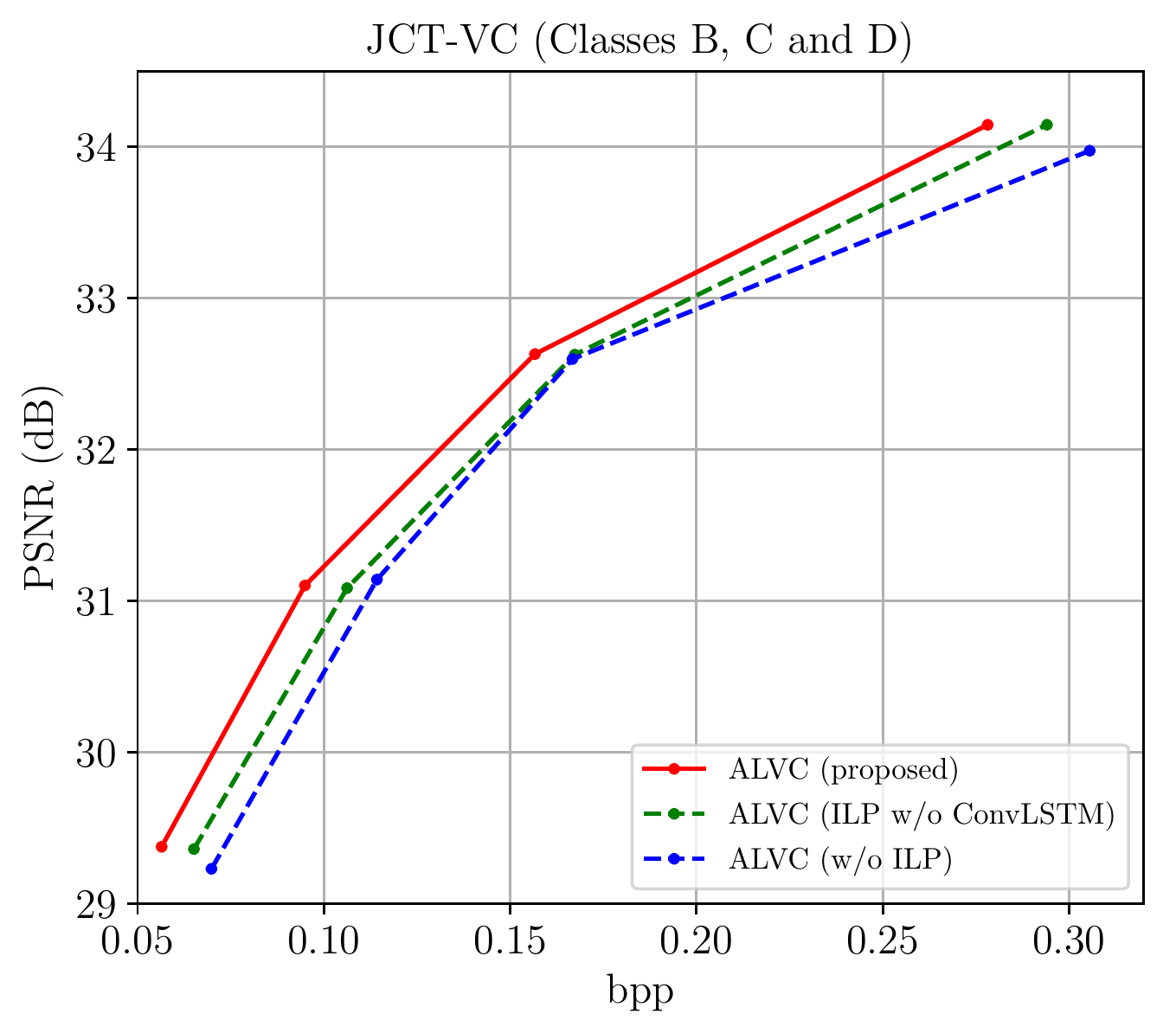}}
\caption{Ablation results of the proposed in-loop prediction.} \label{fig:ilf}
\end{figure*}

\section{Ablation studies}\label{abl}

\subsection{Effectiveness of in-loop prediction}\label{inloop}

We conduct ablation experiments on the JCT-VC dataset (Classes B, C and D) to verify the effectiveness of the proposed in-loop prediction. We train an ALVC model without the in-loop prediction module, which is denoted as ALVC~(w/o~ILP). In ALVC~(w/o~ILP), the reference frame $\hat F_{t-1}$ (or $\hat F_B$ and $\hat F_C$ for B-frames) is directly fed to the location error detector, and therefore, the location error compression module becomes a motion compression module. This way, the framework of ALVC~(w/o~ILP) is similar to most previous works that utilize motion compensation to reduce temporal redundancy. 

\textbf{Prediction quality.} We first evaluate the quality of the predicted frame in ALVC. On B-frames, the average PSNR between the predicted frame $\tilde F_T$ and the target frame $F_T$ are 27.24 dB, 28.08 dB, 28.67 dB and 29.11 dB for $\lambda$ = 256, 512, 1024 and 2048, respectively. While the prediction quality of $\tilde F_t$ on P-frames are 26.78 dB, 27.69 dB, 28.33 dB and 28.72dB, respectively. These values are much higher than the PSNR between the reference frame and the target frame in ALVC~(w/o~ILP), i.e., 22.09 dB, 22.28 dB, 22.40 dB and 22.45 dB, respectively. These results validate the effectiveness of the proposed in-loop prediction. Moreover, the above results also show that the prediction quality on B-frames is obviously better than P-frames. This also verifies the accuracy of our quadratic optical flow prediction strategy in Section~\ref{C}.

\textbf{Location error vs. temporal motion.} Moreover, Fig.~\ref{fig:ilf}-(a) visualizes the location error in ALVC, i.e., the coordinate difference between $\tilde F_t$ and $F_t$, and the temporal motion in ALVC~(w/o~ILP), i.e., the coordinate difference between $\hat F_{t-1}$ (or $\hat F_B$ for B-frame) and $F_t$ (or $F_T$ for B-frame). It can be seen from Fig.~\ref{fig:ilf}-(a) that the location error in ALVC is much smaller than the temporal motion in ALVC~(w/o~ILP) for both P-frames and B-frames, since the proposed in-loop prediction network effectively predicts the target frame. This results in less bit-rate on compressing location error in ALVC than that on compressing motion in ALVC~(w/o~ILP). For example, at $\lambda=2048$, the bit-rate for compressing location error is 0.041 bpp, which is $\sim 20\%$ less than the bit-rate for motion compression (0.051 bpp) in ALVC~(w/o~ILP), and meanwhile, ALVC (34.15 dB) has higher PSNR than ALVC~(w/o~ILP) (33.97 dB). Recall that the proposed in-loop prediction does not consume any more bit-rate.
% The proposed in-loop prediction does not consume any bit-rate, but provides an accurate predicted frame as reference. Therefore, it benefits the compression performance.
The rate-distortion curves of ALVC and ALVC~(w/o~ILP) are illustrated in Fig.~\ref{fig:ilf}-(b), which shows that our in-loop prediction significantly improves the rate-distortion performance. 

{\textbf{Recurrent vs. non-recurrent prediction.}
We also studied into the impact of the recurrent structure in the proposed prediction network for P-frames (Fig.~\ref{fig:2}). We define an ablation model ALVC~(ILP w/o ConvLSTM) that disables the ConvLSTM layers in the prediction network shown in Fig.~\ref{fig:2}. This way, ALVC~(ILP w/o ConvLSTM) uses a non-recurrent frame prediction network for P-frames. The rate-distortion curve of ALVC~(ILP w/o ConvLSTM) is illustrated in Fig.~\ref{fig:ilf}-(b). It can be seen from Fig.~\ref{fig:ilf}-(b) that ALVC (ILP w/o ConvLSTM) performs obviously worse than the proposed ALVC model. This verifies that the recurrent structure plays an important role in capturing long-term temporal information and therefore benefits frame prediction and compression performance. Besides, we can also see that ALVC (ILP w/o ConvLSTM) is still better than the ALVC (w/o ILP), which does not have the in-loop frame prediction networks.
}

\subsection{Analyses on B-frames}

\begin{figure*}[!t]
\centering
\subfigure[B-frame (quadratic vs. linear) and  P-frame]{\includegraphics[width=.32\linewidth]{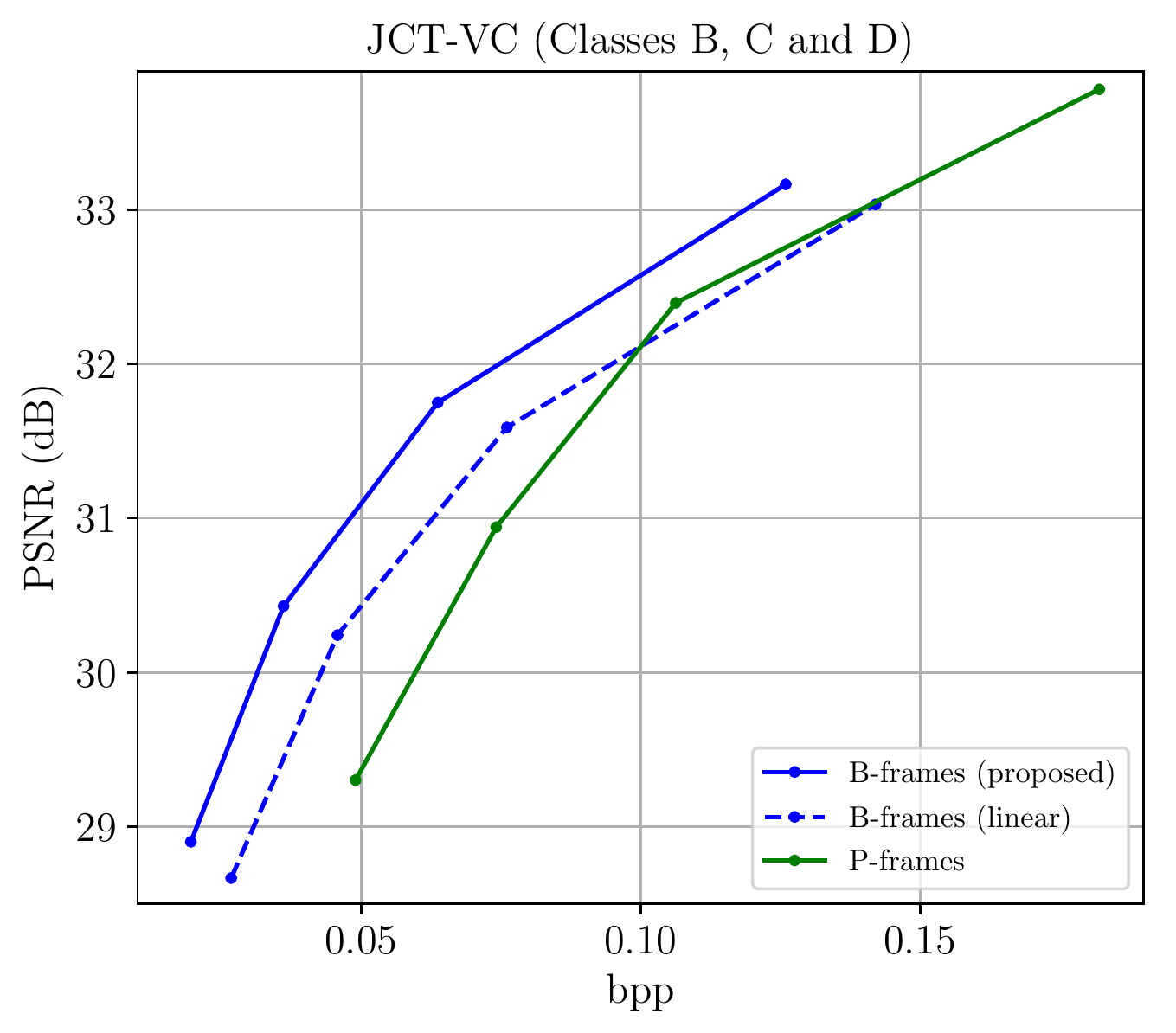}}
\subfigure[Different GOP sizes]{\includegraphics[width=.325\linewidth]{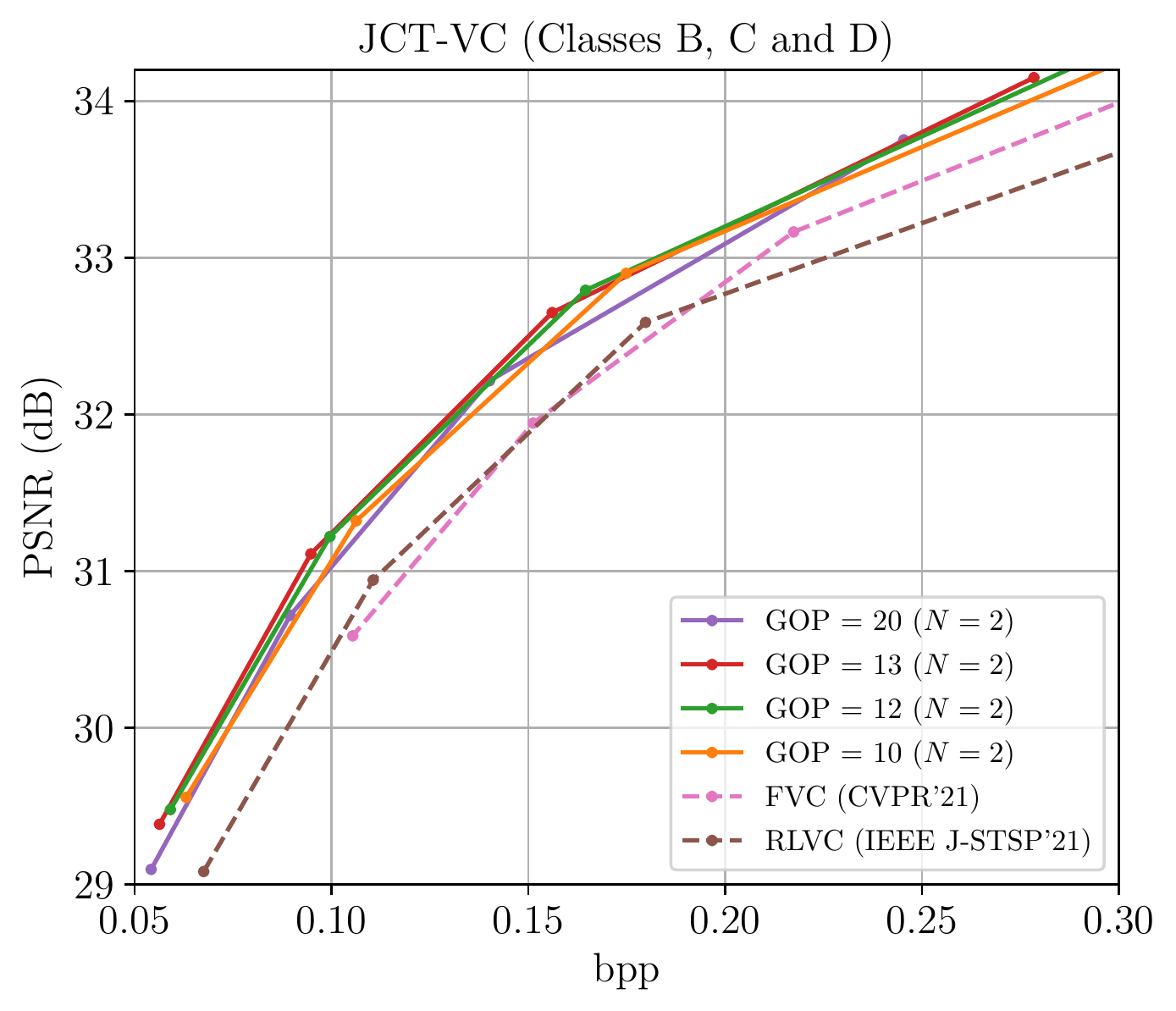}}
\subfigure[Number of B-frames and IPPP mode ]{\includegraphics[width=.325\linewidth]{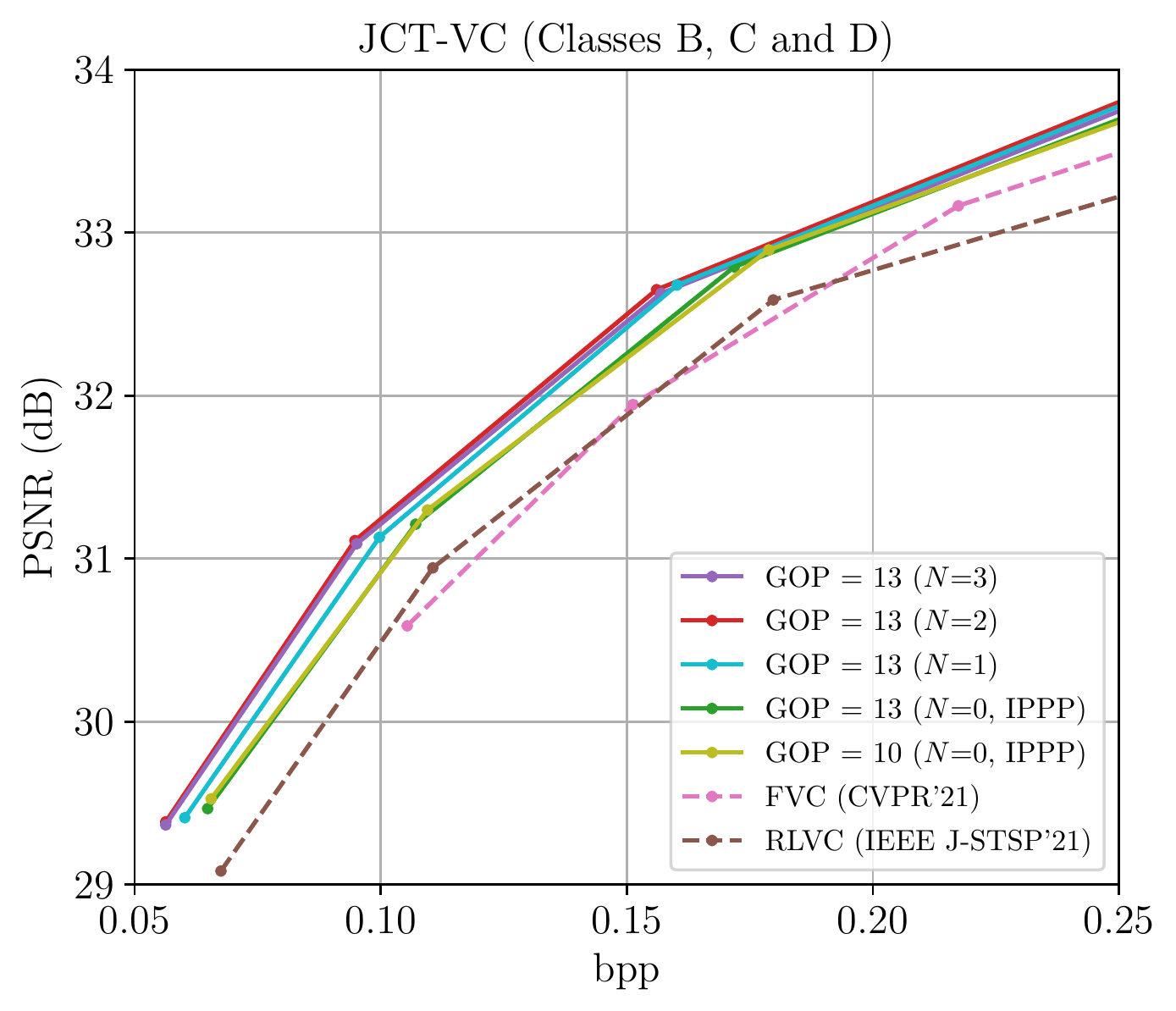}}
\caption{Ablation results of B-frames, P-frames and GOP structures.} \label{fig:gop}
\end{figure*}

\textbf{B-frames vs. P-frames.} As discussed in Section~\ref{inloop}, the proposed in-loop
prediction in B-frames has better prediction quality than in P-frames. As a result,
Fig.~\ref{fig:gop}-(a) shows that the B-frames achieve better compression performance than
P-frames in our ALVC approach. 

\textbf{Quadratic vs. linear prediction in B-frames.} Recall that in the proposed bi-directional in-loop prediction for B-frames, we model the temporal motion in the quadratic order to calculate $f_{B\to T}$ and $f_{C\to T}$ (Section~\ref{C}). In this ablation study, we compare the performance of the quadratic model and linear model. Especially, in the linear model, the temporal movement is modeled as 
\begin{equation}\label{line}
f = v\cdot \Delta t.
\end{equation}
This way, $f_{B\to T}$ and $f_{C\to T}$ are calculated as:
\begin{equation}\label{ll}
f_{B\to T} = f_{B\to C} \cdot \frac{\Delta t_{B\to T}}{\Delta t_{B\to C}}, \quad
f_{C\to T} = f_{C\to B} \cdot \frac{\Delta t_{C\to T}}{\Delta t_{C\to B}},
\end{equation}
where the definitions of $f$ and $\Delta t$ are the same as Section~\ref{C}. Fig.~\ref{fig:gop}-(a) shows the compression performance on B-frames for the proposed (quadratic) model and the linear model in \eqref{ll}. It can be seen that the proposed model obviously improves the performance, indicating the effectiveness of the proposed scheme in Section~\ref{C} for calculating motions on B-frames.

\subsection{GOP structures}

% \textbf{B-frame vs. P-frame.} As discussed above, the proposed in-loop prediction in B-frames has better prediction quality than in P-frames. Therefore, as shown in Fig.~\ref{fig:gop}-(a), the B-frames achieve better compression performance than P-frames in our ALVC approach.

\textbf{GOP size.} Fig.~\ref{fig:gop}-(a) shows the rate-distortion performance of ALVC with different GOP sizes. We first change the GOP to 10 and 12, which are frequently used in previous learned video compression approaches, and then we also enlarge the GOP to 20. As we can see from Fig.~\ref{fig:gop}-(a), the performance is comparable for different GOP sizes, and we outperform the latest methods FVC~\cite{hu2021fvc} and RLVC~\cite{yang2020recurrent} for all. This shows that ALVC is able to work with various GOP sizes, including the large GOPs, such as $\text{GOP} = 20$.

\textbf{Number of consecutive B-frames.} 
In the default setting of our approach, there are two consecutive B-frames ($N=2$) in one GOP. In this experiment, we also test ALVC with one B-frame ($N=1$), three consecutive B-frames ($N=3$) and without B-frame ($N=0$, IPPP) in each GOP. We can see from Fig.~\ref{fig:gop}-(c) that inserting one B-frame ($N=1$) increases the rate-distortion performance from the IPPP mode ($N=0$), and employing two B-frames ($N=2$) in each GOP further boosts the performance. When we insert three consecutive B-frames between two neighboring GOPs, the performance is comparable with $N=2$. This is probably because the increase of $N$ leads to a longer distance between $\hat F_T$ and $\hat F_C$/$\hat F_D$ (refer to Fig.~\ref{fig:pp}), making it hard to further improve the compression performance. Nevertheless, our ALVC with all the above settings achieves better performance than the latest methods FVC and RLVC. This shows that ALVC is able to work with various GOP structures, while maintaining good performance.

\textbf{ALVC (IPPP) vs. FVC and RLVC.} The latest learned video compression approaches FVC~\cite{hu2021fvc} and RLVC~\cite{yang2020recurrent} compress video in the IPPP mode with GOP $=$ 10 and 13, respectively. For a fair comparison, we compare the IPPP mode of ALVC with FVC and RLVC with the same GOP sizes in Fig.~\ref{fig:gop}-(c). 
It can be seen from Fig.~\ref{fig:gop}-(c) that our ALVC with $\text{GOP}=13$ ($N=0$, IPPP) has better performance than RLVC, and our ALVC with $\text{GOP}=10$ ($N=0$, IPPP) also significantly outperforms FVC. These results validate that ALVC is able to beat FVC~\cite{hu2021fvc} and RLVC~\cite{yang2020recurrent} under the same GOP structures. 

\subsection{Impact of I-frame codec}

\begin{figure}[!t]
\centering
\includegraphics[width=.8\linewidth]{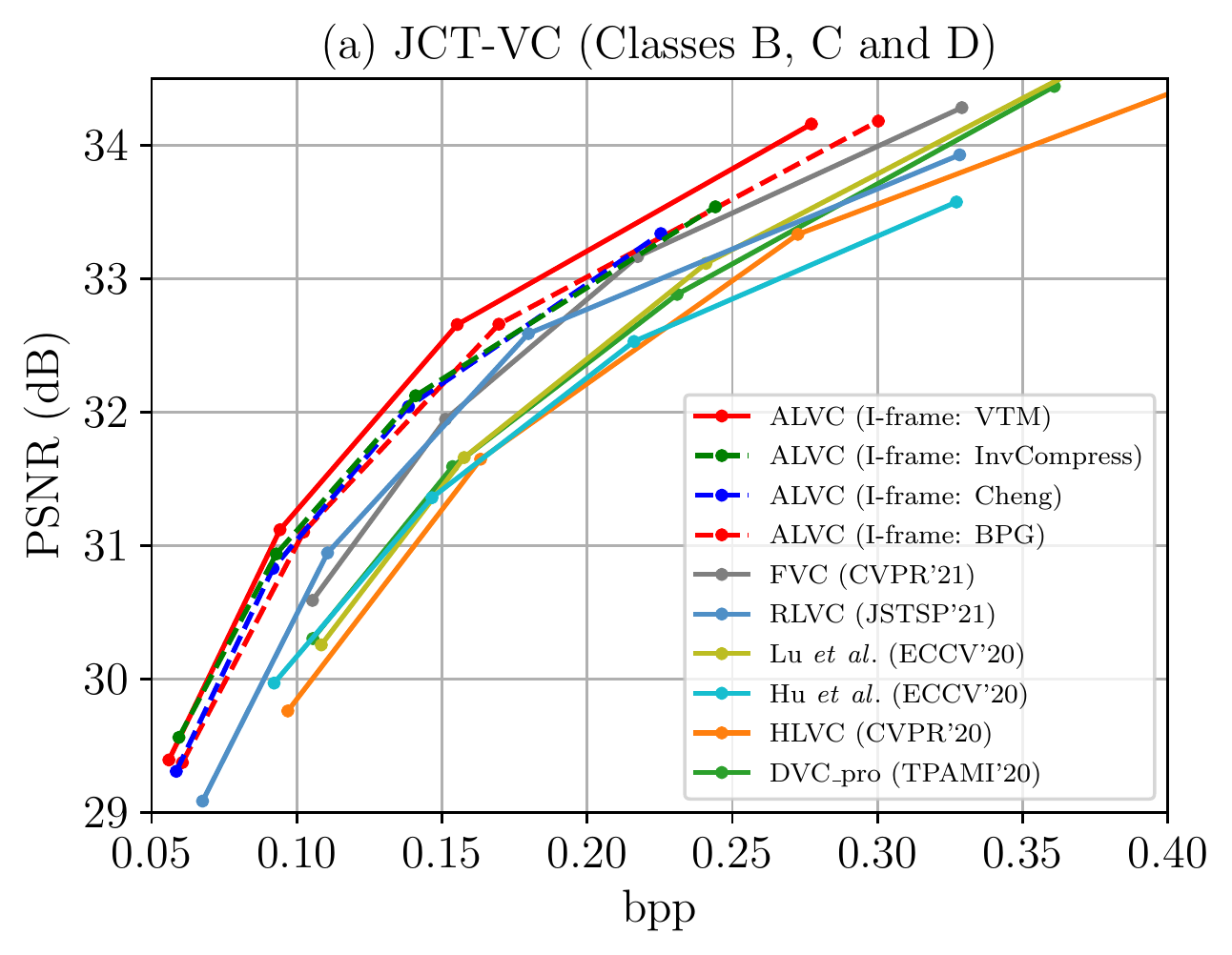}
\caption{Performance of ALVC with I-frames compressed by HEVC (BPG), Cheng et al. \cite{cheng2020learned} and Xie et al. \cite{xie2021enhanced}, which are shown in dash lines.} \label{fig:I}
\end{figure}

In the PSNR model of ALVC, we use VTM~\cite{VTM} to compress I-frames. In the latest learned video compression approaches FVC~\cite{hu2021fvc} and RLVC~\cite{yang2020recurrent}, I-frames are compressed by HEVC (BPG~\cite{BPG}). For a fairer comparison, we illustrate the performance of the proposed ALVC with I-frames compressed by HEVC (BPG) in Fig.~\ref{fig:I}. We can see from Fig.~\ref{fig:I} that when using HEVC (BPG) on I-frames in our ALVC model (dash line), we still outperform FVC and RLVC, and achieve the state-of-the-art performance in learned video compression methods. 

{Moreover, to make our PSNR model fully learned, i.e., without handcrafted codec, we further replace the VTM with the end-to-end learned image compression approaches~\cite{cheng2020learned,xie2021enhanced} to compress I-frames, which are defined as ALVC (I-frame: Cheng) and ALVC (I-frame: InvCompress), respectively. It can be seen from Fig.~\ref{fig:I} that ALVC (I-frame: Cheng) and ALVC (I-frame: InvCompress) both outperform other learned video compression approaches. That is, the proposed ALVC approach is able to achieve state-of-the-art performance with fully learned frameworks.}

\section{Conclusion and future work}
This paper proposes the novel in-loop prediction modules for advancing learned video compression. Such a module learns to predict the target frame from the historical prior in the previously compressed frames without consuming any bit-rate, and only the prediction error needs to be encoded into bitstreams. Given an effective prediction, the bit-rate can be significantly reduced. The proposed method shows better performance than the existing learned compression approaches, and also beats the default setting of x265 on PSNR and the SSIM-tuned slowest setting of x265 on MS-SSIM. The ablation studies show the effectiveness of the proposed in-loop prediction, and verify that ALVC is able to adjust to different GOP sizes. Besides, the IPPP mode of ALVC and the ALVC with I-frames compressed by HEVC (BPG) also outperform the last learned compression methods FVC and RLVC.

{In this paper, the proposed approach is optimized towards distortion, i.e., PSNR and MS-SSIM, and therefore the compressed frames may suffer from over-smoothness. Employing a discriminator in ALVC to train it with an adversarial loss is probably a way to address this issue. For example, most recently, the GAN-based perceptual video compression approach~\cite{yang2021perceptual} has been proposed. On the one hand, the discriminator in~\cite{yang2021perceptual} is possible to be utilized to the proposed ALVC network, and this way, ALVC can be optimized by the GAN-loss to generate compressed frames with sharp and photo-realistic textures. On the other hand, due to the advanced performance of ALVC, it can serve as a better generator for~\cite{yang2021perceptual} to further advance its perceptual performance. These can be seen as interesting future works.}

\bibliographystyle{IEEEtran}
\bibliography{bare_jrnl}

\begin{IEEEbiography}[{\includegraphics[width=1\linewidth]{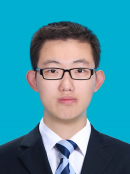}}]{Ren Yang} is a doctoral student at ETH Zurich, Switzerland. He received his M.Sc. degree in 2019 at the School of Electronic and Information Engineering, Beihang University, China, and his B.Sc. degree at the same university in 2016. His research interests include computer vision and video compression. He has published several papers in top international journals and conferences, such as IEEE T-PAMI, IEEE T-IP, IEEE T-CSVT, CVPR, IJCAI, ICCV and ICME. He is a Senior Program Committee (SPC) Member of IJCAI 2021, and he serves as a reviewer for CVPR, ICCV, ECCV, NeurIPS, ICLR, IEEE T-IP, IEEE J-STSP, IEEE T-CSVT, IEEE T-MM, Elsevier's SPIC and NEUCOM, etc. He is the Winner of the Three Minute Thesis (3MT) Competition at IEEE ICME 2019. He is also a co-organizer of the NTIRE 2021, NTIRE 2022 and AIM 2022 Workshops and Challenges, and a co-organizer/speaker of the Tutorials in ACM MM 2021, CVPR 2021 and IEEE VCIP 2020.
\end{IEEEbiography}

\begin{IEEEbiography}[{\includegraphics[width=1\linewidth]{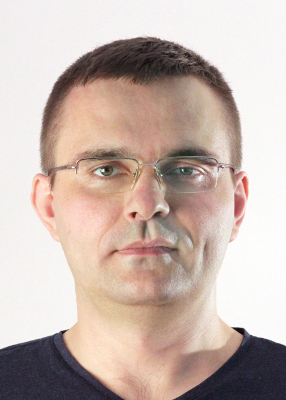}}]{Radu Timofte} 
received his PhD degree in Electrical Engineering from the KU Leuven, Belgium, in 2013. Currently, he is a professor and holds the Chair for Computer Science IV (Computer Vision) at the University of W\"urzburg, Germany. Also, he is a lecturer and a group leader at ETH Z\"urich, Switzerland. He is a member of the editorial board of top journals such as IEEE TPAMI, Elsevier's CVIU and NEUCOM, and SIAM's SIIMS. He regularly serves as an area chair and as a reviewer for top conferences such as CVPR, ICCV, IJCAI, and ECCV. His work received several awards. Radu Timofte is the 2022 awardee of an Alexander von Humboldt Professorship for Artificial Intelligence. He is a co-founder of Merantix and a co-organizer of NTIRE, CLIC, AIM, Mobile AI and PIRM workshops and challenges. His current research interests include deep learning, mobile AI, visual tracking, computational photography, image/video compression, restoration, enhancement and manipulation.
\end{IEEEbiography}

\begin{IEEEbiography}[{\includegraphics[width=1\linewidth]{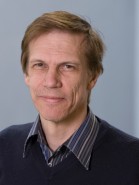}}]{Luc Van Gool}
received the degree in electromechanical engineering at the Katholieke Universiteit Leuven, in 1981. Currently, he is a professor at the Katholieke Universiteit Leuven in Belgium and the ETH Zurich in Switzerland. He leads computer vision research at both places, where he also teaches computer vision. He has been a program committee member of several major computer vision conferences. His main interests include 3D reconstruction and modeling, object recognition, tracking, and gesture analysis, and the combination of those. He received several Best Paper awards and was nominated Distinguished Researcher by the IEEE Computer Science committee. He received a David Marr Prize, and a Koenderink and a U.V. Helava award. He is a co-founder of 10 spin-off companies. He is a member of the IEEE.
\end{IEEEbiography}

\clearpage
\onecolumn
\begin{center}
\LARGE
Advancing Learned Video Compression with In-loop Frame Prediction
\vspace{.5em}

-- Supporting Document --

\end{center}

\vspace{1em}

\begin{center}
Ren Yang, Radu Timofte, Luc Van Gool
\end{center}

\section*{Appendix}
\subsection{Detailed network architecture}

\begin{figure*}[!h]
\centering
\includegraphics[width=.8\linewidth]{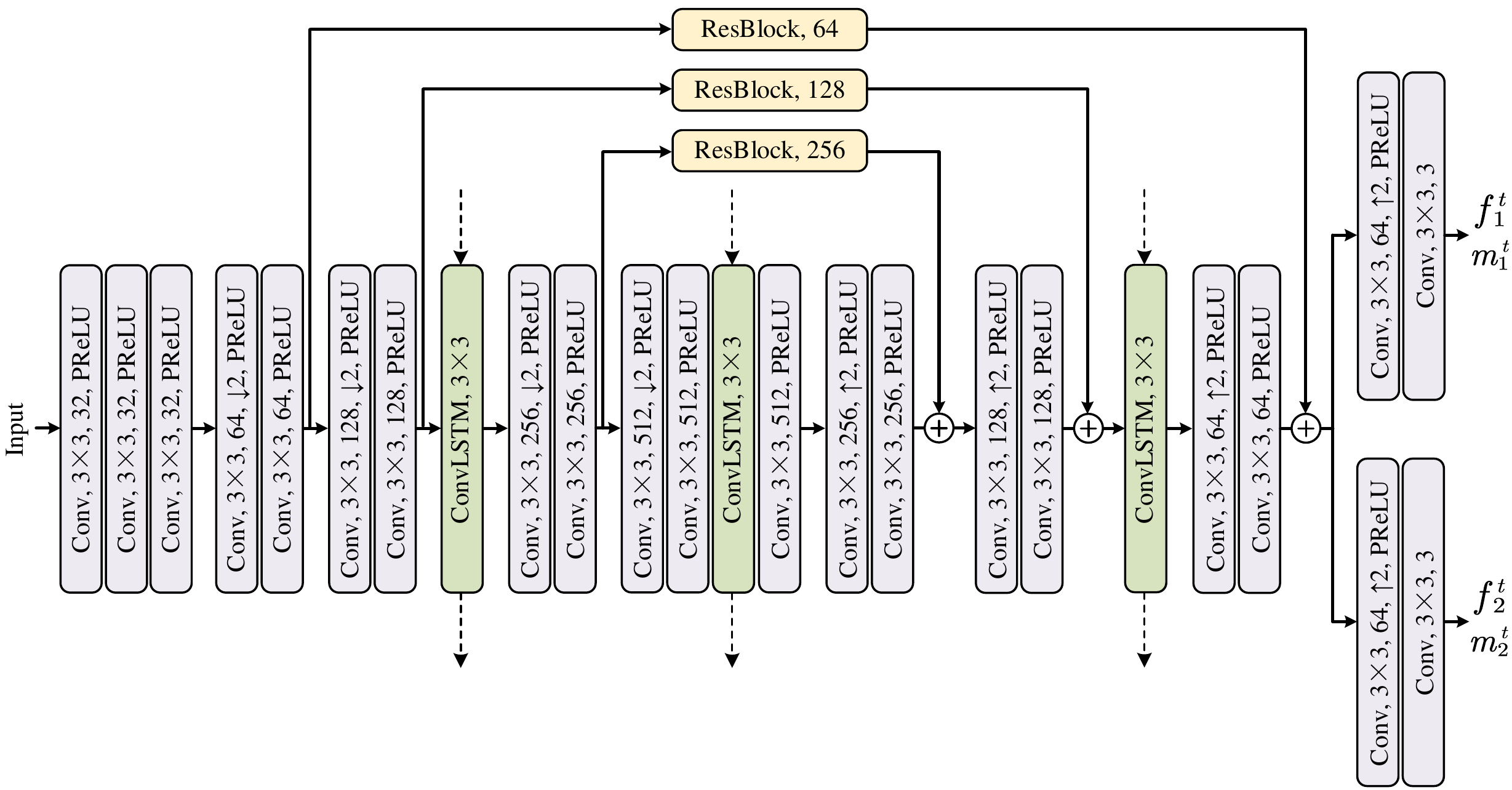}
\caption{Architecture of the RecPred network in Fig.~\ref{fig:2}. The dash lines indicate the hidden states transferred through the recurrent cells.} \label{fig:recpred}
\end{figure*}

\begin{figure*}[!h]
\centering
\includegraphics[width=.8\linewidth]{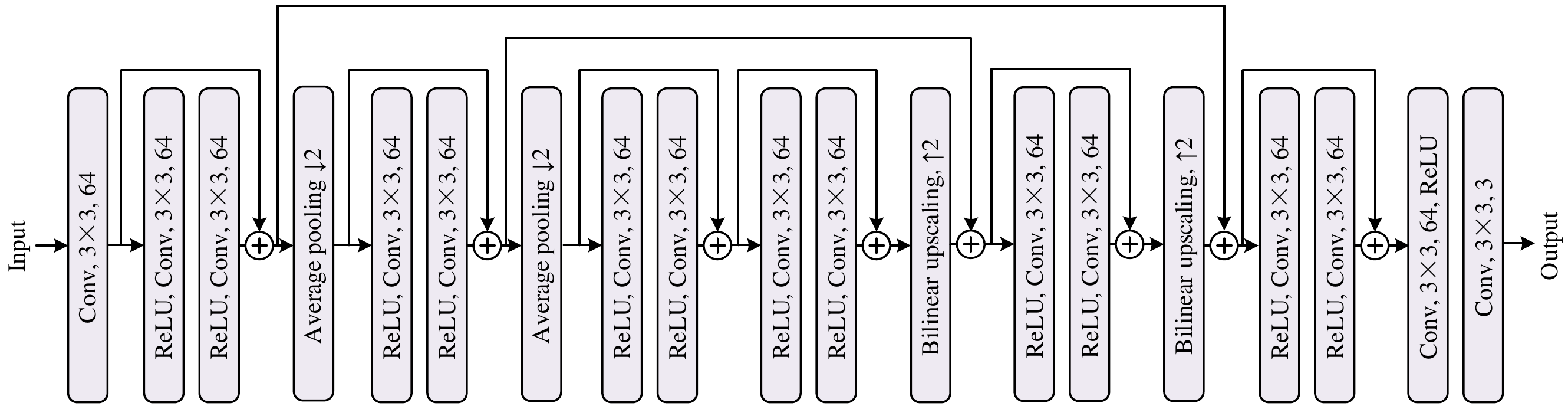}
\caption{We use the network with the same architecture as the motion compensation network in DVC~\cite{lu2019dvc}, HLVC~\cite{yang2020learning} and RLVC~\cite{yang2020recurrent} for the merging ($M$) operation and location error correction in ALVC. ``ReLU'' before ``Conv'' denotes the pre-activation convolutional layer.} \label{fig:merge}
\end{figure*}

In the proposed ALVC model for P-frames (refer to Fig.~\ref{fig:2}), the detailed architecture of the proposed RecPred network is illustrated in Fig.~\ref{fig:recpred}. Besides, the network for merging ($M$) and the network of location error correction (after warping) share the same architecture, which is shown in Fig.~\ref{fig:merge}. In these figures, the convolutional layers are denoted as ``Conv, filter size, filter number'', and we use $\downarrow2$ and $\uparrow2$ to denote $\times 2$ downscaling and $\times 2$ upscaling, respectively. ``PReLU'' indicates the parametric rectified linear unit with a learnable parameter.
The ``ResBlock, filter number'' in Fig.~\ref{fig:recpred} indicates a ResBlock with two layers of ``Conv, $3\times 3$, filter number, PReLU'' and a skip connection.

\begin{figure*}[!h]
\centering
\includegraphics[width=.8\linewidth]{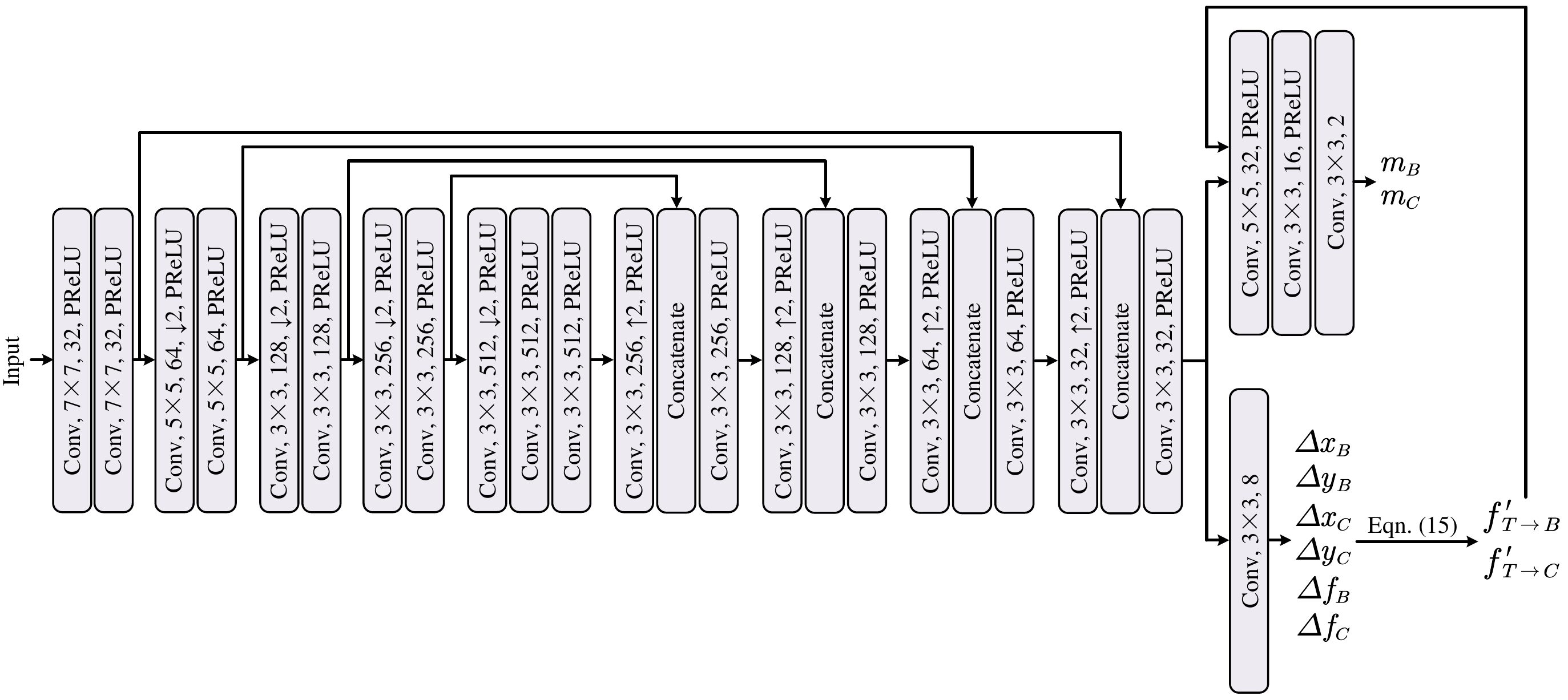}
\caption{Architecture of the RefineNet in Fig.~\ref{fig:bi}, which is based on the adaptive flow filtering network in \cite{xu2019quadratic}.} \label{fig:refine}
\end{figure*}

In the ALVC model for B-frames (refer to Fig.~\ref{fig:bi}), the merging network and the location error correction network are the same as those for P-frames. Fig.~\ref{fig:refine} (on the next page) shows the architecture of the RefineNet in the model for B-frames. In Fig.~\ref{fig:refine}, the outputs $\Delta x_B$, $\Delta x_C$, $\Delta y_B$, $\Delta y_C$, $\Delta f_B$ and $\Delta f_C$ are utilized to refine the optical flows $f_{T\to B}$ and $f_{T\to C}$. Specifically, the refined flows are obtained as:
\begin{equation}
\begin{split}
f'_{T\to B}(x,y) &= f_{T\to B}(x+\Delta x_B, y+\Delta y_B) + \Delta f_B, \\
f'_{T\to C}(x,y) &= f_{T\to C}(x+\Delta x_C, y+\Delta y_C) + \Delta f_C,
\end{split}
\end{equation}
Then, the refine flows $f'_{T\to B}$ and $f'_{T\to C}$ and the masks $m_B$ and $m_C$ are fed into the next procedures in Fig.~\ref{fig:bi}.


\newpage
\subsection{Per-sequence rate-distortion curves}

\begin{figure}[!t]
\centering
\includegraphics[width=\linewidth]{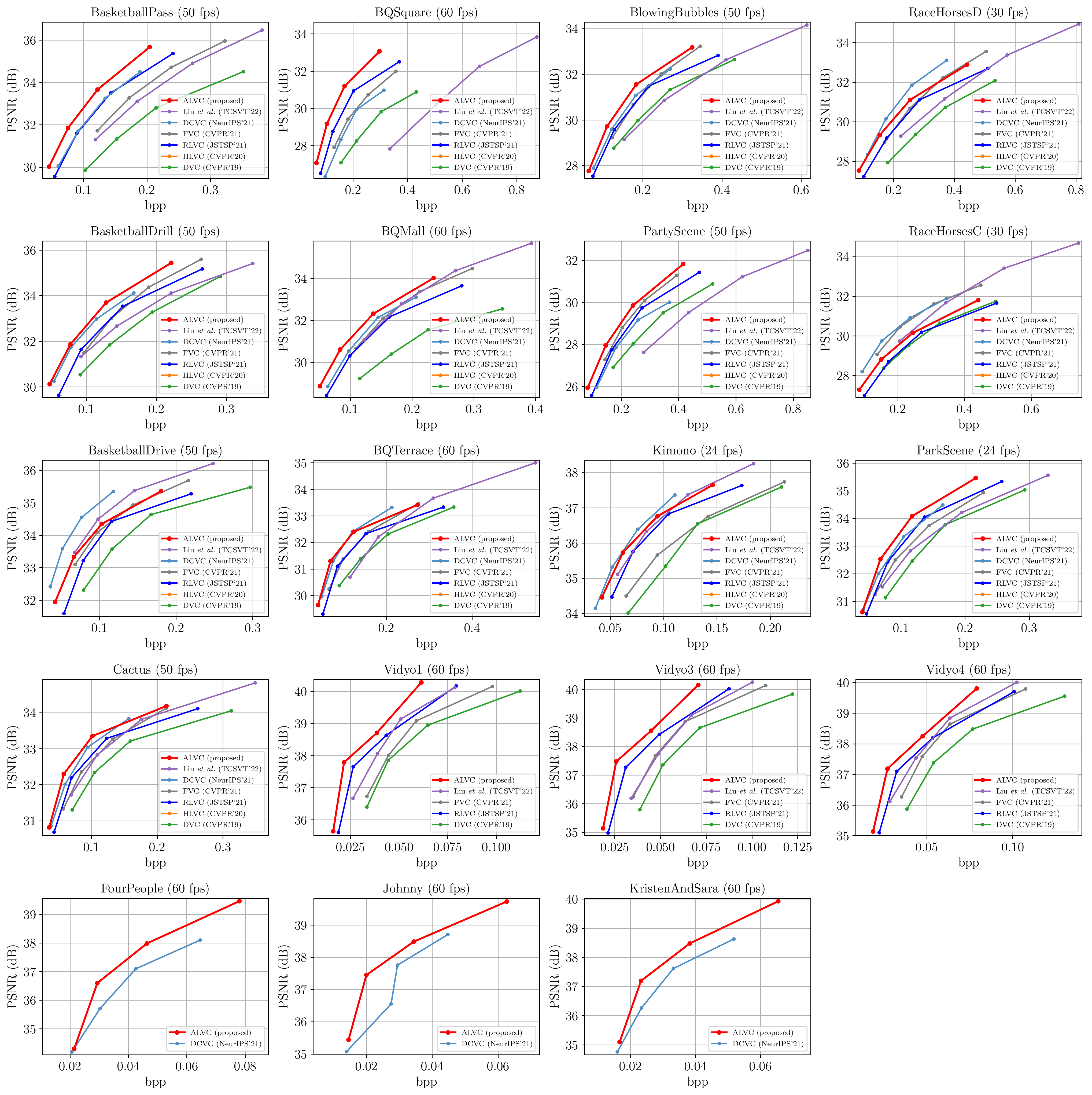}
\caption{The rate-distortion curve (PSNR) of every sequence in JCT-VC Classes B, C, D, E and E'.} \label{fig:rd_all}
\end{figure}

\begin{figure}[!t]
\centering
\includegraphics[width=\linewidth]{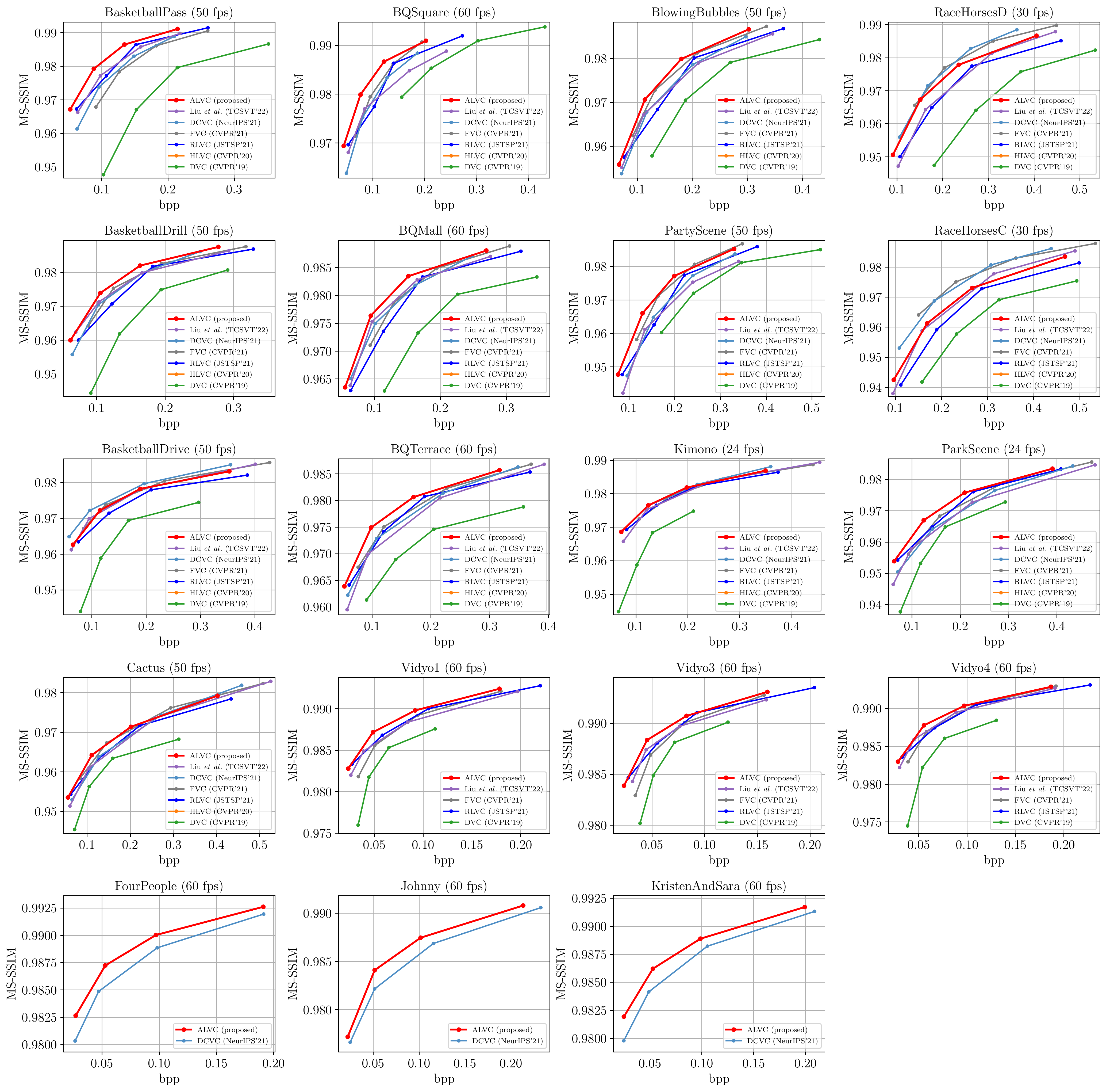}
\caption{The rate-distortion curve (MS-SSIM) of every sequence in JCT-VC Classes B, C, D, E and E'.} \label{fig:rd_all_ssim}
\end{figure}

We plot the rate-distortion curve of every sequence in JCT-VC Classes B, C, D, E and E' in terms of PSNR and MS-SSIM in Figure~\ref{fig:rd_all} and Figure~\ref{fig:rd_all_ssim}, respectively. Please see the next page. These figures show that we achieve state-of-the-art performance on most videos in the test sets.

\end{document}